\documentclass[twocolumn,tighten,twocolappendix]{aastex63}

\usepackage{amssymb,amsmath,amsthm}
\usepackage{graphicx}
\usepackage{mathrsfs}
\usepackage{booktabs}
\usepackage{multirow}
\usepackage{empheq}
\usepackage{verbatim}
\usepackage{mathtools}
\usepackage{hyperref}
\usepackage{xcolor}
\usepackage{xspace}
\usepackage{lineno}
\usepackage[makeroom]{cancel}
\usepackage[acronym]{glossaries}

\def\uv{\ensuremath{(u,v)}\xspace} 
\def\uas{\ensuremath{\mu}as\xspace} 
\def\am{\textit{am}\xspace} 
\def\ngehtsim{\texttt{ngehtsim}\xspace} 
\def\m87{M87$^*$\xspace} 
\def\sgra{Sgr\,A$^*$\xspace} 

\defcitealias{M87Paper1}{EHTC et al. (2019a)}
\defcitealias{M87Paper2}{EHTC et al. (2019b)}
\defcitealias{M87Paper3}{EHTC et al. (2019c)}
\defcitealias{M87Paper4}{EHTC et al. (2019d)}
\defcitealias{M87Paper5}{EHTC et al. (2019e)}
\defcitealias{M87Paper6}{EHTC et al. (2019f)}
\defcitealias{M87Paper7}{\m87~Paper~7}
\defcitealias{M87Paper8}{\m87~Paper~8}
\defcitealias{SgrA_Paper1}{\sgra~Paper~1}
\defcitealias{SgrA_Paper2}{\sgra~Paper~2}
\defcitealias{SgrA_Paper3}{\sgra~Paper~3}
\defcitealias{SgrA_Paper4}{\sgra~Paper~4}
\defcitealias{SgrA_Paper5}{\sgra~Paper~5}
\defcitealias{SgrA_Paper6}{\sgra~Paper~6}

\begin{document}

\title{Atmospheric limitations for high-frequency ground-based VLBI}

\correspondingauthor{Dominic~W.~Pesce}
\email{dpesce@cfa.harvard.edu}

\author[0000-0002-5278-9221]{Dominic~W.~Pesce}
\affiliation{Center for Astrophysics $|$ Harvard \& Smithsonian, 60 Garden Street, Cambridge, MA 02138, USA}
\affiliation{Black Hole Initiative at Harvard University, 20 Garden Street, Cambridge, MA 02138, USA}

\author[0000-0002-9030-642X]{Lindy~Blackburn}
\affiliation{Center for Astrophysics $|$ Harvard \& Smithsonian, 60 Garden Street, Cambridge, MA 02138, USA}
\affiliation{Black Hole Initiative at Harvard University, 20 Garden Street, Cambridge, MA 02138, USA}

\author{Ryan~Chaves}
\affiliation{Center for Astrophysics $|$ Harvard \& Smithsonian, 60 Garden Street, Cambridge, MA 02138, USA}

\author{Sheperd~S.~Doeleman}
\affiliation{Center for Astrophysics $|$ Harvard \& Smithsonian, 60 Garden Street, Cambridge, MA 02138, USA}
\affiliation{Black Hole Initiative at Harvard University, 20 Garden Street, Cambridge, MA 02138, USA}

\author{Mark~Freeman}
\affiliation{Center for Astrophysics $|$ Harvard \& Smithsonian, 60 Garden Street, Cambridge, MA 02138, USA}

\author[0000-0002-5297-921X]{Sara~Issaoun}
\affiliation{Center for Astrophysics $|$ Harvard \& Smithsonian, 60 Garden Street, Cambridge, MA 02138, USA}
\affiliation{Black Hole Initiative at Harvard University, 20 Garden Street, Cambridge, MA 02138, USA}

\author[0000-0002-4120-3029]{Michael~D.~Johnson}
\affiliation{Center for Astrophysics $|$ Harvard \& Smithsonian, 60 Garden Street, Cambridge, MA 02138, USA}
\affiliation{Black Hole Initiative at Harvard University, 20 Garden Street, Cambridge, MA 02138, USA}

\author[0000-0002-6100-4772]{Greg~Lindahl}
\affiliation{Common Crawl Foundation, 9663 Santa Monica Blvd. \#425, Beverly Hills, CA 90210, USA}
\affiliation{Center for Astrophysics $|$ Harvard \& Smithsonian, 60 Garden Street, Cambridge, MA 02138, USA}

\author[0000-0001-8242-4373]{Iniyan~Natarajan}
\affiliation{Center for Astrophysics $|$ Harvard \& Smithsonian, 60 Garden Street, Cambridge, MA 02138, USA}
\affiliation{Black Hole Initiative at Harvard University, 20 Garden Street, Cambridge, MA 02138, USA}

\author[0000-0003-4622-5857]{Scott~N.~Paine}
\affiliation{Center for Astrophysics $|$ Harvard \& Smithsonian, 60 Garden Street, Cambridge, MA 02138, USA}

\author[0000-0002-7179-3816]{Daniel~C.~M.~Palumbo}
\affiliation{Center for Astrophysics $|$ Harvard \& Smithsonian, 60 Garden Street, Cambridge, MA 02138, USA}
\affiliation{Black Hole Initiative at Harvard University, 20 Garden Street, Cambridge, MA 02138, USA}

\author[0000-0001-5461-3687]{Freek~Roelofs}
\affiliation{Center for Astrophysics $|$ Harvard \& Smithsonian, 60 Garden Street, Cambridge, MA 02138, USA}
\affiliation{Black Hole Initiative at Harvard University, 20 Garden Street, Cambridge, MA 02138, USA}

\author[0000-0003-3826-5648]{Paul~Tiede}
\affiliation{Center for Astrophysics $|$ Harvard \& Smithsonian, 60 Garden Street, Cambridge, MA 02138, USA}
\affiliation{Black Hole Initiative at Harvard University, 20 Garden Street, Cambridge, MA 02138, USA}

\begin{abstract}
Very long baseline interferometry (VLBI) provides the highest-resolution images in astronomy.  The sharpest resolution is nominally achieved at the highest frequencies, but as the observing frequency increases so too does the atmospheric contribution to the system noise, degrading the sensitivity of the array and hampering detection.  In this paper, we explore the limits of high-frequency VLBI observations using \ngehtsim, a new tool for generating realistic synthetic data. \ngehtsim uses detailed historical atmospheric models to simulate observing conditions, and it employs heuristic visibility detection criteria that emulate single- and multi-frequency VLBI calibration strategies. We demonstrate the fidelity of \ngehtsim's predictions using a comparison with existing 230\,GHz data taken by the Event Horizon Telescope (EHT), and we simulate the expected performance of EHT observations at 345\,GHz.  Though the EHT achieves a nearly 100\% detection rate at 230\,GHz, our simulations indicate that it should expect substantially poorer performance at 345\,GHz; in particular, observations of \m87 at 345\,GHz are predicted to achieve detection rates of $\lesssim$20\% that may preclude imaging.  Increasing the array sensitivity through wider bandwidths and/or longer integration times -- as enabled through, e.g., the simultaneous multi-frequency upgrades envisioned for the next-generation EHT -- can improve the 345\,GHz prospects and yield detection levels that are comparable to those at 230\,GHz. \m87 and \sgra observations carried out in the atmospheric window around 460\,GHz could expect to regularly achieve multiple detections on long baselines, but analogous observations at 690 and 875\,GHz consistently obtain almost no detections at all.
\end{abstract}

\section{Introduction}

The history of very long baseline interferometry (VLBI) has seen a progression towards observations at ever-higher frequencies.  Beginning with a series of early results spanning frequencies lower than 1\,GHz up to 22\,GHz \citep{Broten_1967,Bare_1967,Moran_1967,Burke_1970}, important milestones have included the first VLBI detections at frequencies of 43\,GHz \citep{Moran_1979}, 89\,GHz \citep{Readhead_1983}, and 223\,GHz \citep{Padin_1990}.  The Event Horizon Telescope (EHT) currently executes the highest-frequency VLBI observations of any existing array, and EHT observations at a frequency of $\sim$230\,GHz have demonstrated the unique science opportunities accessible to an instrument capable of imaging with an angular resolution of $\sim$20\,\uas.  EHT images of \m87 \citep{M87Paper1,M87Paper2,M87Paper3,M87Paper4,M87Paper5,M87Paper6,M87Paper7,M87Paper8,M87Paper9} and \sgra \citep{SgrA_Paper1,SgrA_Paper2,SgrA_Paper3,SgrA_Paper4,SgrA_Paper5,SgrA_Paper6} currently provide our only event-horizon-scale views of the emission region in the immediate vicinity of a black hole.

It is notable that the amount of time separating the first successful VLBI observations at any wavelength (conducted in 1967) and the first VLBI detections at a wavelength of $\sim$1.3\,mm (for which the observations were conducted in 1989) was shorter than the amount of time separating the first 1.3\,mm detections and the first 1.3\,mm images (for which the observations were conducted in 2017).  This seeming discrepancy provides some indirect evidence for the many practical difficulties facing high-frequency VLBI observations, which must contend with smaller typical aperture sizes, higher characteristic system noise temperatures, increased atmospheric attenuation, and shorter atmospheric coherence times than analogous observations conducted at lower frequencies.  The decades preceding the first EHT results saw substantial developments in broadband instrumentation, with the explicit aim of overcoming these difficulties through improvements to the instantaneous sensitivity of the array \citep{Doeleman_2008,Whitney_2013,Vertatschitsch_2015,Matthews_2018,M87Paper2}.

Difficulties notwithstanding, the allure of high-frequency VLBI observations continues to motivate developments that push the frontier.  The EHT has already carried out VLBI observations at 345\,GHz \citep{Raymond_2024,Crew_2023}, and the next-generation EHT project seeks to substantially build out the array \citep{Doeleman_2019,Doeleman_2023}: increasing the bandwidth, placing multiple new dishes at locations that fill gaps in the existing coverage, and adopting simultaneous multi-frequency observing capabilities that aim to make 345\,GHz observations more commonplace.  Furthermore, there are aspirations for at least a subset of the stations to carry out VLBI at frequencies as high as 690\,GHz \citep[e.g.,][]{Inoue_2014}.  Maximizing the potential success of such endeavors, as well as making decisions about the optimal locations of new dishes and about which technological developments to prioritize, requires an informed understanding of how array upgrades and operational choices will manifest in the collected data.

A number of synthetic data generation tools have been developed and used within the high-frequency VLBI community to address such considerations.  In addition to predicting future capabilities, synthetic data are important for developing imaging and calibration algorithms \citep[e.g.,][]{Roelofs_2023} and even for analyzing existing data \citep[e.g.,][]{M87Paper4,SgrA_Paper3}.  Two synthetic data tools that have seen substantial use within the EHT collaboration are \texttt{ehtim} \citep{Chael_2016,Chael_2018,Chael_2023} and \texttt{SYMBA} \citep{Roelofs_2020}.  \texttt{ehtim} is foremost an image reconstruction software, but it also includes a suite of easy-to-use synthetic observation and data handling abilities. \texttt{ehtim} provides rapid (typically $\sim$seconds) data generation with a simple user interface. However, it does not directly simulate many effects (particularly those associated with the atmosphere) that are relevant for determining array sensitivity.  In contrast, \texttt{SYMBA} was developed for end-to-end synthetic data generation from the start; it leverages \texttt{MeqSilhouette} \citep{Blecher_2017,Natarajan_2022} to simulate physically realistic atmospheric and instrumental effects, and it passes all data through the \texttt{rPICARD} VLBI calibration pipeline \citep{Janssen_2019}.  By mirroring the same processes that are applied to real VLBI data, \texttt{SYMBA} aims to maximize realism at the cost of increased runtime (typically tens of minutes to hours, depending on the generated data volume) and requisite user sophistication relative to \texttt{ehtim}.

In this paper we introduce \ngehtsim\footnote{\url{https://github.com/Smithsonian/ngehtsim}}, a Python-based tool for simulating radio interferometric data that builds on the work of \citet{Raymond_2021} to incorporate realistic atmospheric conditions and visibility detection criteria.  \ngehtsim aims to retain the speed and user-friendliness of synthetic data generation tools like \texttt{ehtim} while striving for the physical realism of tools like \texttt{SYMBA}.  Relevant atmospheric properties are pre-tabulated and stored within the \ngehtsim codebase, which serves both as a centralized repository of weather information and also reduces computational overhead during synthetic data generation.  \ngehtsim also employs heuristic visibility detection schemes that emulate VLBI calibration strategies relevant for high-frequency and multi-frequency observations without requiring the running of a full calibration pipeline.

This paper is organized as follows.  In \autoref{sec:Atmosphere} we describe our procedure for generating atmospheric spectra, and in \autoref{sec:SyntheticData} we describe how the synthetic interferometric data are generated.  In \autoref{sec:Examples} we provide several examples of \ngehtsim synthetic data generation, focusing primarily on EHT observations at frequencies of 230\,GHz and above.  We summarize and conclude in \autoref{sec:Summary}.

\section{Simulating atmospheric conditions} \label{sec:Atmosphere}

The primary atmospheric quantities that determine the sensitivity of radio astronomical observations are the optical depth ($\tau$) and the brightness temperature ($T_b$).  From the optical depth, the atmospheric transmittance is given by $e^{-\tau}$ and quantifies the amount of incident radiation that gets absorbed or scattered by the atmosphere.  We use version 12.2 of the \am radiative transfer software \citep{Paine_2022} to compute atmospheric optical depth and brightness temperature as a function of frequency ($\nu$).

\begin{figure}
    \centering
    \includegraphics[width=1.00\columnwidth]{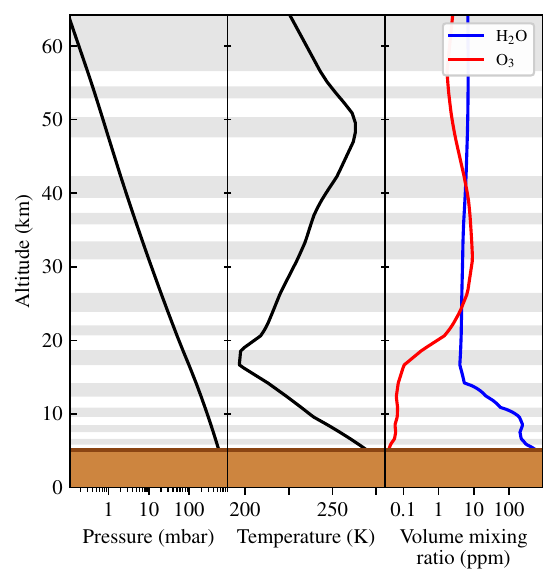}
    \caption{Atmospheric state above the ALMA site at $(\text{latitude},\text{longitude}) = (-23.032^{\circ},-67.755^{\circ})$ and elevation of 5040\,meters, taken from the April 15, 2022 MERRA-2 reanalysis at 0 UTC.  From left to right, the quantities plotted are the pressure in millibar (mbar), the temperature in K, and the volume mixing ratio of H$_2$O (in blue) and O$_3$ (in red) in parts per million (ppm); all quantities are plotted as a function of altitude.  The alternating white and gray shading indicates the locations of the individual atmospheric layers, and the brown shaded region on the bottom indicates the ground level.}\label{fig:atmospheric_triplet}
\end{figure}

\subsection{Weather data and atmospheric spectra} \label{sec:MERRA}

The \am software takes as its primary inputs the pressure, temperature, and composition at each of a user-defined number of layers in the atmosphere.  We set the temperature of the radiation that is incident onto the upper atmospheric layer equal to the CMB temperature, $T_{\text{CMB}} = 2.725$\,K \citep{Mather_1999,Fixsen_2009}.

We obtain atmospheric state information from NASA's Modern-Era Retrospective analysis for Research and Applications version 2 (MERRA-2) database, which assimilates past measurements into a general circulation model to produce an estimate of the daily meteorological history of Earth's atmosphere dating back to the year 1980 \citep{Rienecker_2011,Molod_2015,Gelaro_2017}.  Among the quantities available in the MERRA-2 database are the temperature ($T$), specific humidity ($q$), mass mixing ratio of liquid water ($w_{\text{LWP}}$), mass mixing ratio of ice water ($w_{\text{IWP}}$), mass mixing ratio of ozone (O$_3$; $w_{\text{O$_3$}}$), and the wind speed in the eastward and northward directions, all as a function of the pressure coordinate ($P$).  Each of these quantities is regridded from native model coordinates onto latitude (every 0.5\,degrees), longitude (every 0.625\,degrees), pressure altitude (on up to 42 standard levels), and time (every 3\,hours UTC).  An example set of atmospheric quantities is shown in \autoref{fig:atmospheric_triplet}.

The \am code requires mixing ratios of H$_2$O and O$_3$ to be specified in terms of volume rather than mass.  The H$_2$O mass mixing ratio $w_{m,\text{H$_2$O}}$ is related to the specific humidity $q$ by

\begin{equation}
w_{m,\text{H$_2$O}} = \frac{q}{1-q} . \label{eqn:WaterMassMixingRatio}
\end{equation}

\noindent To convert from the mass mixing ratio ($w_{m,\text{H$_2$O}}$) to the volume mixing ratio ($w_{\text{H$_2$O}}$), we multiply by the ratio of the relative molecular masses of air and water,

\begin{equation}
w_{\text{H$_2$O}} = 1.6078 \, w_{m,\text{H$_2$O}} . \label{eqn:WaterVolumeMixingRatio}
\end{equation}

\noindent We apply a similar conversion to the O$_3$ mass mixing ratio provided in the MERRA-2 database, for which the conversion factor is 0.6034.

The liquid water path (LWP) and ice water path (IWP) are provided by MERRA-2 as mass mixing ratios ($w$) in each pressure layer, but \am requires column densities ($S$).  We convert from the former to the latter for each atmospheric layer using

\begin{equation}
S = \frac{w \Delta P}{g} , \label{eqn:ColumnDensityConversion}
\end{equation}

\noindent where $\Delta P$ is the pressure difference between the upper and lower boundaries of the atmospheric layer and $g$ is the standard acceleration of gravity on the surface of the Earth.

The \am software also requires a choice of frequency resolution ($\Delta \nu$) and frequency range $(\nu_{\text{min}},\nu_{\text{max}})$ over which to compute optical depths and brightness temperatures.  Unless otherwise specified, we use values of $\nu_{\text{min}} = 0$\,THz, $\nu_{\text{max}} = 2$\,THz, and $\Delta \nu = 1$\,GHz for all computations in this paper (see \autoref{app:Resolution}).

\subsection{Interpolation scheme} \label{sec:Interpolation}

\begin{figure}
    \centering
    \includegraphics[width=1.00\columnwidth]{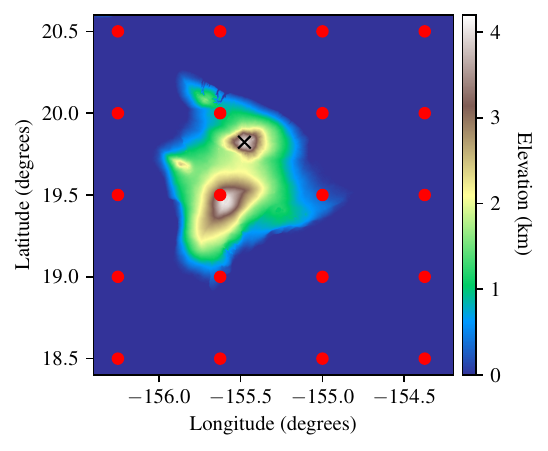}
    \caption{Topographical map of the island of Hawaii.  The map is colored by elevation, and the overlaid red circles indicate the locations of the MERRA-2 grid points.  The black ``$\times$'' marks the summit of Mauna Kea, which is the location of the submillimeter observing facilities JCMT and SMA.  The MERRA-2 database provides only a few grid points covering the entire island, necessitating interpolation to estimate the atmospheric information above a specific telescope site (see \autoref{sec:Interpolation}).}\label{fig:map_MaunaKea}
\end{figure}

The MERRA-2 data are supplied on a coarse grid in latitude and longitude, with individual grid cells measuring several tens of kilometers on a side.  This cell size is much larger than the size of a typical mountain or other site on which a telescope might be located (see \autoref{fig:map_MaunaKea}), so we interpolate the MERRA-2 values when determining the atmospheric properties for a particular site.

For each site, we first identify the four MERRA-2 grid points that enclose it.  All atmospheric quantities at each of these four locations are linearly interpolated in elevation onto a log-uniform grid of 100 atmospheric layers that span an elevation range $[z_{\text{site}},\text{70\,km}]$, where $z_{\text{site}}$ is the elevation of the site itself and 70\,km is chosen because it is roughly equal to the maximum elevation modeled by MERRA-2.  The values of the elevation-dependent quantities in each layer are then bilinearly interpolated from the four surrounding grid point locations to the location of the site itself.

\subsection{Weather tabulation}

Using the interpolated atmospheric state information from the MERRA-2 database as inputs to \am, we have tabulated a variety of ``weather parameters'' for more than 80 sites that host existing or near-future radio or (sub)millimeter facilities around the globe.  A partial list of sites is provided in \autoref{app:SiteTable}, and the complete list can be accessed from within \ngehtsim.  For each of these sites, we have tabulated the following weather parameters:

\begin{itemize}
    \item The ground-level air pressure ($P$), which we determine from the values recorded in the MERRA-2 database by interpolating to the location and elevation of the site as described in \autoref{sec:Interpolation}.
    \item The ground-level air temperature ($T_{\text{air}}$), which we similarly determine from the MERRA-2 data after appropriate interpolation.
    \item The ground-level wind speed ($v_{\text{wind}}$), which we determine as the quadrature sum of the eastward and northward wind speeds recorded in the MERRA-2 database, after interpolating each to the location and elevation of the site.
    \item The zenith PWV, which is computed within \am as an integral over the total water vapor column above the site.
    \item The zenith atmospheric optical depth ($\tau_z$) as a function of frequency, which is computed within \am and recorded in \ngehtsim on a frequency range that spans $[0,2]$\,THz with a 1\,GHz spacing.  To minimize the data volume, we use a principal component analysis (PCA) decomposition to compress the optical depth spectra stored in \ngehtsim; our specific compression scheme is detailed in \autoref{app:Decomposition}.
    \item The zenith atmospheric brightness temperature ($T_{b,z}$) as a function of frequency, which is also computed within \am and is recorded and compressed in an analogous manner to the optical depth.
\end{itemize}

\begin{figure*}
    \centering
    \includegraphics[width=1.00\textwidth]{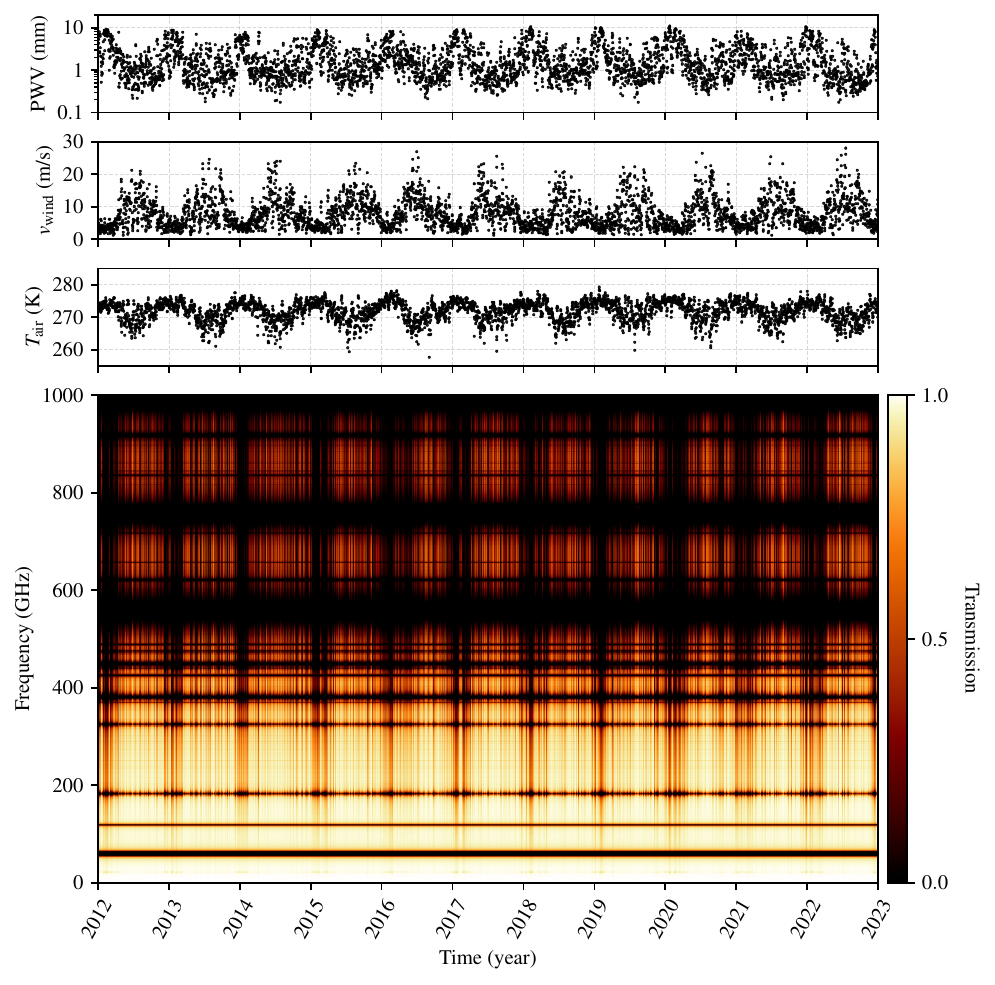}
    \caption{Example weather information tabulated in \ngehtsim, shown for the ALMA site.  The top panel shows the zenith PWV versus time, the second panel shows the ground-level wind speed ($v_{\text{wind}}$) versus time, the third panel shows the ground-level air temperature ($T_{\text{air}}$) versus time, and the bottom panel shows the zenith atmospheric transmission ($e^{-\tau_z}$) as a function of both time and frequency.}\label{fig:weather_vs_time_and_frequency}
\end{figure*}

The MERRA-2 database provides atmospheric state information every 3 hours, but we average each of the above weather parameters on a per-day basis for the purposes of tabulation within \ngehtsim (see \autoref{app:ResolutionInTime}).  These weather parameters have been calculated for all dates from 2012 January 1 up to 2023 January 1, and all are available as precomputed data tables within the \ngehtsim package. An example time series of some of the weather parameters that can be accessed within \ngehtsim is shown in \autoref{fig:weather_vs_time_and_frequency}.

\section{Simulating interferometric data} \label{sec:SyntheticData}

The weather parameters tabulated within \ngehtsim are used to determine telescope sensitivities during synthetic data generation.  In this section we detail how \ngehtsim operates.

\subsection{Initial data generation} \label{sec:DataGeneration}

Given a source model and a choice of array configuration, \ngehtsim uses the \texttt{ehtim} library \citep{Chael_2016,Chael_2018,Chael_2023} to generate \uv-coverage corresponding to a synthetic observation.  We start by determining the values of the ``Stokes visibilities'' at each \uv point, which are initially set equal to the Fourier transform of the source model per \citep{Thompson_2017}

\begin{equation}
\tilde{I}(u,v) = \iint I(x,y) e^{2 \pi i (ux + vy)} dx dy . \label{eqn:FT}
\end{equation}

\noindent Here, $I$ is the source model Stokes I brightness as a function of location $(x,y)$ in the image plane, and $\tilde{I}$ represents the Stokes I visibilities.  An expression analogous to \autoref{eqn:FT} is also used to determine the initial Stokes Q, U, and V visibilities from the corresponding source model brightness maps.  These Stokes visibilities are then converted to a circular correlation product representation via

\begin{equation}
\begin{pmatrix}
V_{RR} \\
V_{RL} \\
V_{LR} \\
V_{LL}
\end{pmatrix} = \begin{pmatrix}
\tilde{I} + \tilde{V} \\
\tilde{Q} + i \tilde{U} \\
\tilde{Q} - i \tilde{U} \\
\tilde{I} - \tilde{V}
\end{pmatrix} .
\end{equation}

\noindent Currently, \ngehtsim only produces data in a circular polarization basis, which is appropriate for most existing VLBI arrays.  We assume that all visibilities have had absolute flux density calibration applied, such that both the measurements and their uncertainties can be expressed in physical units (e.g., Jy) rather than dimensionless correlation coefficients and signal-to-noise ratios.  Gaussian random systematic errors to amplitude calibration are optionally applied.

\subsection{Baseline sensitivity} \label{sec:BaselineSensitivity}

The sensitivity of the baseline comprised of stations $i$ and $j$ is characterized by a thermal noise level, $\sigma_{ij}$, determined by the radiometer equation and can be expressed as

\begin{equation}
\sigma_{ij} = \frac{1}{\eta_q} \sqrt{\frac{\text{SEFD}^*_i \text{SEFD}^*_j}{2 \Delta \nu \Delta t}} . \label{eqn:ThermalNoise}
\end{equation}

\noindent Here, $\Delta \nu$ is the frequency bandwidth over which the measurement is being integrated for a single correlation product, $\Delta t$ is the corresponding integration time,

\begin{equation}
\text{SEFD}^* = \frac{2 k T_{\text{sys}}}{\eta_{\text{ff}} \eta_{w} A_{\text{eff}}} e^{\tau} \label{eqn:SEFD}
\end{equation}

\noindent is the opacity-corrected ``system equivalent flux density'' (SEFD) for a station with system temperature $T_{\text{sys}}$ and effective collecting area $A_{\text{eff}}$, $k$ is the Boltzmann constant, $\tau$ is the line-of-sight atmospheric optical depth, $\eta_{\text{ff}}$ is the forward efficiency of the antenna \citep[controlling the degree of spillover; see, e.g.,][]{Mangum_2002}, $\eta_w$ is an efficiency factor associated with the wind buffeting the telescope, and $\eta_q$ is an efficiency factor associated with digitization of the signal.  Throughout this paper we assume $\eta_q = 0.88$, appropriate for 2-bit quantization \citep{Thompson_2017}.  The thermal noise level given by \autoref{eqn:ThermalNoise} is assumed to be the same for all four correlation products on a single baseline.

\subsubsection{System temperature}

For each site, the system temperature is given by

\begin{equation}
T_{\text{sys}} = \left[ T_{\text{rx}} + \eta_{\text{ff}} T_{b,\text{inc}} + \left( 1 - \eta_{\text{ff}} \right) T_{\text{gnd}} \right] \left( 1 + r \right) , \label{eqn:SystemTemperature}
\end{equation}

\noindent where $T_{\text{rx}}$ is the receiver temperature, $T_{b,\text{inc}}$ is the brightness temperature of the radiation incident on the dish, $T_{\text{gnd}}$ is the ground temperature, and $r$ is the sideband separation ratio (defined such that a perfectly sideband-separating receiver has $r=0$ and a perfect double-sideband receiver has $r=1$).  Receiver temperatures differ from site to site and across frequencies, so \ngehtsim includes some default values but also permits users to specify their own receiver temperatures.  We set $T_{\text{gnd}} = T_{\text{air}}$ from the MERRA-2 database.  An accurate determination of $\eta_{\text{ff}}$ would require an elevation-dependent integral over the antenna beam pattern for each telescope \citep[see, e.g.,][]{Rusch_1970}, but for source elevations above $\sim$20\,degrees a typical amount of spillover contributes to the system temperature at the $\sim$several percent level \citep[e.g.,][]{Potter_1973,Greve_1998,Kramer_2013,Mangum_2017}.  For the simulations in this paper, we assume a value of $\eta_{\text{ff}} = 0.95$ for all antennas.

The incident brightness temperature $T_{b,\text{inc}}$ contains contributions from the atmosphere, the CMB, and the source itself, and it is given by

\begin{equation}
T_{b,\text{inc}} = T_{\text{atm}} \left( 1 - e^{-\tau} \right) + \left( T_{\text{CMB}} + T_{\text{source}} \right) e^{-\tau} . \label{eqn:BrightnessTemp}
\end{equation}

\noindent Here, $T_{\text{atm}}$ is the effective atmospheric temperature, 

\begin{equation}
T_{\text{source}} = \frac{F_{\text{tot}} A_{\text{eff}}}{2k} \label{eqn:SourceBrightnessTemp}
\end{equation}

\noindent is the brightness temperature of the source, and $F_{\text{tot}}$ is the total flux density of the source.  We use a plane-parallel atmosphere approximation to obtain $\tau$ from the zenith optical depth $\tau_z$ via

\begin{equation}
\tau = \frac{\tau_z}{\sin(\theta_{\text{el}})} ,
\end{equation}

\noindent where $\theta_{\text{el}}$ is the elevation angle of the observed source.  $\tau_z$ is computed using \am as described in \autoref{sec:Atmosphere}.

Because the atmosphere does not have a single temperature, \am instead computes a zenith atmospheric brightness temperature ($T_{b,z}$) that integrates over contributions from the full column of atmosphere above a site.  To include an elevation dependence, we first determine an effective atmospheric temperature $T_{\text{atm}}$ using

\begin{equation}
T_{\text{atm}} = \frac{T_{b,z} - T_{\text{CMB}} e^{-\tau_z}}{1 - e^{-\tau_z}} . \label{eqn:EffectiveTatm}
\end{equation}

\noindent The effective atmospheric temperature is then appropriately scaled for non-zenith elevations when computing the system temperature (see \autoref{eqn:BrightnessTemp}).

\subsubsection{Effective area}

The effective collecting area $A_{\text{eff}}$ of a single-dish site is given by

\begin{equation}
A_{\text{eff}} = \frac{\pi D^2 \eta_{\text{ap}}}{4} ,
\end{equation}

\noindent where $\eta_{\text{ap}}$ is the aperture efficiency and $D$ is the dish diameter.  For phased-array sites, we use an effective total diameter of

\begin{equation}
D = \sqrt{\frac{4}{\pi} \sum_j A_{j}} ,
\end{equation}

\noindent where $A_{j}$ is the geometric area of the $j$th dish in the array and the sum is taken over all dishes in the array.

The aperture efficiency is determined by Ruze's law \citep{Ruze_1952,Ruze_1966},

\begin{equation}
\eta_{\text{ap}} = \exp\left[ - \left( \frac{4 \pi \sqrt{\sigma_{\text{RMS}}^2 + \sigma_{\text{off}}^2}}{\lambda} \right)^2 \right]
\end{equation}

\noindent where $\sigma_{\text{RMS}}$ is the RMS surface accuracy of the dish, $\sigma_{\text{off}}$ is the typical focus offset in equivalent units of surface accuracy, and $\lambda$ is the observing wavelength.  Effective dish diameters and RMS surface accuracies for existing and near-future sites whose weather information is tabulated in \ngehtsim are provided in \autoref{app:SiteTable}.  Throughout this paper we assume $\sigma_{\text{off}} = 10$\,$\mu$m, which is larger than the magnitude of defocus measured for some dishes \citep[e.g., ALMA; see][]{Mangum_2006} but still subdominant to the surface accuracy limitations for all antennas simulated within \ngehtsim.

\subsubsection{Sensitivity degradation from wind}

High wind speeds can cause issues with telescope pointing, and intermittent changes in wind speed can rock a dish back and forth, decreasing its average sensitivity.  We use the wind speeds tabulated from the MERRA-2 database to derive a phenomenological efficiency factor, $\eta_w$, associated with these wind-related sensitivity losses.

We use a logistic function to capture the wind efficiency as a function of wind speed $v_{\text{wind}}$,

\begin{equation}
\eta_w = 1 - \frac{1}{1 + \exp\left[ -\frac{w}{2} \left( \frac{v_{\text{wind}}}{2(v_s - v_d)} - 1 \right) \right]} .
\end{equation}

\noindent Here, $v_d$ and $v_s$ set the wind speeds within which substantial degradation occurs, such that for $v_{\text{wind}} \lesssim v_d$ there is very little degradation and for $v_{\text{wind}} \gtrsim v_s$ the degradation is substantial; the quantity $w$ sets the values of $\eta_w$ at $v = v_d$ and $v = v_s$.  Throughout this paper, we assume values of $v_d = 15$\,m\,s$^{-1}$, $v_s = 25$\,m\,s$^{-1}$, and $w = 10$.  Given these settings, the wind efficiency takes on values of $\eta_w \approx 0.8$ when $v_{\text{wind}} = v_d$ and $\eta_w \approx 0.2$ when $v_{\text{wind}} = v_s$.  While we expect that the actual degree of wind-loading for an antenna will depend in detail on the antenna structure, observing frequency, and source elevation, the selected values are similar to specifications for telescopes such as ALMA \citep{Greve_2008} and the GLT \citep{Raffin_2014}.

\subsection{Visibility detection scheme} \label{sec:VisDetection}

Given a known source flux density (\autoref{sec:DataGeneration}) and sensitivity (\autoref{sec:BaselineSensitivity}) for each baseline, \ngehtsim uses a multi-step visibility detection scheme that seeks to emulate the process of fringe-finding.  As is evident in \autoref{eqn:ThermalNoise}, a baseline can in principle achieve arbitrary sensitivities by simply increasing the integration time $\Delta t$.  However, rapid visibility phase fluctuations induced by changes in the atmospheric water vapor content over each site prevent coherent integration of visibilities for periods of time that are comparable to or longer than the atmospheric coherence timescale (i.e., the timescale over which the phase variance is equal to 1 radian), which is typically a few tens of seconds for millimeter-wavelength observations.  The primary question that drives visibility detectability is thus whether the source can be detected with sufficient sensitivity, and within a sufficiently short integration time, such that these rapid phase variations can be tracked and removed via calibration.  \autoref{app:AtmosphericPhase} provides a more detailed discussion of the statistical character of these phase fluctuations and how baseline sensitivity and integration time impact our ability to track them.

The visibility detection scheme used in \ngehtsim is based on signal-to-noise ratio (SNR) considerations, where we define the SNR ($\rho$) to be

\begin{equation}
\rho \equiv \frac{1}{\sqrt{2} \sigma} \left( \frac{\left| V_{RR} + V_{LL} \right|}{2} \right) ,
\end{equation}

\noindent as appropriate for the Stokes I signal. Here, $\sigma$ is the thermal noise (see \autoref{eqn:ThermalNoise}) appropriate for this baseline at the observing time and frequency of interest.  Given this definition for $\rho$, the detection algorithm proceeds as follows:

\begin{enumerate}
    \item For each baseline, we first assess whether the SNR is sufficient to track phases on that baseline.  Following \autoref{sec:PhaseTracking}, we consider a ``strong'' baseline to be one that achieves $\rho \geq \rho_{\text{thresh}}$ within an integration time of $\Delta t$.  The visibilities on all strong baselines are considered to be detected.
    \item If simultaneous multi-frequency observations are being simulated, then frequency phase transfer (FPT) can be used to assist visibility detection (see \autoref{sec:PhaseTrackingMultiFreq}).  For each baseline, if the SNR at the reference frequency is at least $R \rho_{\text{thresh}}$ (where $R$ is the frequency ratio between the target and reference frequencies) within an integration time of $\Delta t$, then we consider that baseline to be ``strong'' and the visibility on that baseline is considered to be detected.
    \item Following the approach developed in \citet{Blackburn_2019}, we partition the strong baselines into groups of mutually connected stations.  All baselines between stations within a single such ``fringe group'' are considered to be detectable.  Baselines that connect two stations contained in different fringe groups remain undetected.
\end{enumerate}

\noindent Throughout this paper, we assume a threshold SNR value of $\rho_{\text{thresh}} = 5$ and an integration time equal to one-third of the coherence time\footnote{One-third of the coherence time is the integration time over which $\sim$90\% of the visibility amplitude is recovered.}; atmospheric coherence times are assumed to be (90, 30, 20, 10)\,seconds for observing frequencies of (86, 230, 345, 690)\,GHz. While we expect that coherence times should generically vary with telescope location and local weather conditions, and that they should also evolve with time throughout the duration of a single observation, the values selected here are approximately representative of the atmospheric conditions above millimeter-wavelength sites such as those participating in EHT observations \citep[][see also \autoref{app:AtmosphericPhase}]{M87Paper2}.

If the visibility on a baseline is detected, then we assume that all four correlation products on that baseline can be coherently integrated for arbitrarily long periods of time.  Even so, we note that some of the ``detections'' resulting from the above procedure can still end up having arbitrarily low SNR; such visibilities may be more appropriately treated as consistent with nondetection at the level of the achieved final sensitivity.

\section{Example observations} \label{sec:Examples}

In this section, we provide several examples of synthetic data generated using \ngehtsim for current and potential future high-frequency VLBI observations with the EHT.  All of the simulations presented here use version 1.0.0 of the \ngehtsim software.

\subsection{EHT 2017 observations of \m87} \label{sec:EHT2017}

\begin{figure*}
    \centering
    \includegraphics[width=1.00\textwidth]{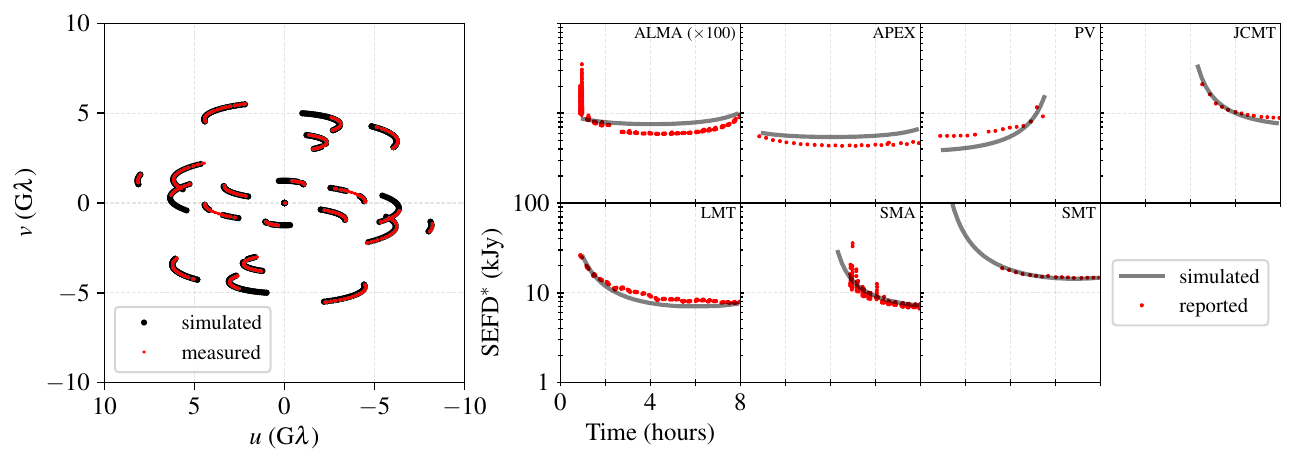}
    \caption{A comparison of the \uv-coverage (left) and reported station SEFDs (right) between the real April 6, 2017 EHT data of \m87 and a simulation of these data generated using \ngehtsim; see \autoref{sec:EHT2017} for simulation details. In all panels, the simulations are plotted in black and the corresponding measurements or reported values are plotted in red.  All of the SEFD panels on the right share common horizontal and vertical axis ranges, which are explicitly labeled in the LMT panel.  Note that the SEFDs for the ALMA station have been increased by a factor of 100 to enable plotting on the same scale as used for the other stations.} \label{fig:EHT2017}
\end{figure*}

The first EHT observing campaign from which images of \m87 and \sgra were produced took place in April of 2017 \citep{M87Paper3}.  Calibrated data for both targets are publicly available online\footnote{\url{https://eventhorizontelescope.org/for-astronomers/data}}, and telescope metadata are also available\footnote{\url{https://github.com/eventhorizontelescope/2020-D02-01}}.  In this section we compare an \m87 dataset from this observing campaign to corresponding data simulated using \ngehtsim.

\subsubsection{Array and observation characteristics}

\begin{deluxetable}{lc}
\tablecolumns{2}
\tablewidth{0pt}
\tablecaption{Simulation parameters for EHT 2017\label{tab:EHT2017}}
\tablehead{\colhead{Parameter} & \colhead{Value(s)}}
\startdata
\multirow{2}{*}{sites} & ALMA, APEX, JCMT, \\
 & LMT, PV, SMA, SMT \\
target source & \m87 \\
frequency & 227.1\,GHz \\
bandwidth & 2\,GHz \\
date & April 6, 2017 \\
starting time & 1\,UT \\
track duration & 7 hours \\
\enddata
\tablecomments{Parameters determining the structure of the observing track for our simulation of the April 6, 2017 EHT observations of \m87.}
\end{deluxetable}

To mimic the real EHT observations, we use the settings specified in \autoref{tab:EHT2017} to define the structure of the observing track.  We aim to simulate the April 6, 2017 observation of \m87 at the ``low-band'' observing frequency of 227.1\,GHz; this observing track began at roughly 1\,UT, lasted for approximately 7 hours, and utilized 2\,GHz of bandwidth for fringe-finding.  We simulate the atmospheric conditions at every site using the procedure described in \autoref{sec:Atmosphere}, starting with the MERRA-2 data for the specific April 6, 2017 date.  The assumed receiver properties for each participating telescope are detailed in \autoref{app:Receivers}, and the dish properties are provided in \autoref{app:SiteTable}.

Though \ngehtsim contains information about the diameter of each telescope in the EHT array, not all dishes were able to use their full collecting area during the 2017 EHT observing campaign.  As detailed in \cite{M87Paper2}, the ALMA array observed with only 37 dishes (corresponding to an effective dish diameter of $\sim$73\,meters) and the LMT was operating with an effective diameter of only 32.5\,meters\footnote{At the time, only 32.5\,meters out of the final 50\,meters of the LMT dish surface had been paneled.}.  We thus override the default dish sizes for these stations in our simulation.  Furthermore, some sites suffered from unmodeled sensitivity losses that are captured as ``multiplicative mitigation factors'' in \cite{M87Paper3}; these factors inflate the system temperature in a manner similar to the sideband separation ratio.  For our simulation, we follow \cite{M87Paper3} and increase the PV system temperature by a factor of 3.663 and the SMA system temperature by a factor of 1.4.

For the source structure we use a geometric model that captures both the gross features observed in the \m87 image \citep{M87Paper4} as well as finer-scale features expected from theory \citep{Johnson_2020}; a detailed description of the source structure model is provided in \autoref{app:SourceModels}.

\subsubsection{Simulation results and comparison with real data}

\autoref{fig:EHT2017} shows the results of the \ngehtsim simulation.  The left panel compares the \uv-coverage obtained from \ngehtsim (in black) with that from the actual EHT observations (in red).  The two sets of coverage are qualitatively similar, with the most notable differences being that some tracks appear to persist for longer in the synthetic data than they do in the real data.  These tracks correspond to baselines containing the SMT station, which lost several scans at the beginning of the April 6, 2017 observing track \citep{M87Paper3}.  The technical issues that resulted in these dropped scans are not simulated by \ngehtsim, which thus overpredicts the amount of data on baselines containing the SMT.

The right panels of \autoref{fig:EHT2017} compare the simulated SEFDs of each telescope (in black) with the a priori estimates (in red) contained in the EHT metadata.  Gross trends in the SEFDs at all telescopes are captured well by the simulation, with systematic deviations typically at the $\sim$10\% level.  Short-lived deviations are evident in the ALMA and SMA SEFDs, which exhibit large spikes in the measurements.  These SEFD spikes are associated with momentary losses of phasing efficiency, as both of these sites join EHT observations as phased arrays; phasing efficiency is not simulated by \ngehtsim, so it underpredicts the SEFDs during periods of poor array phasing.  More systematic offsets are seen for a few stations (ALMA, APEX, and -- most severely -- PV), which likely arise from mismatches between the simplified antenna models assumed within \ngehtsim and the true performance of the antennas.  However, we note that the final SEFDs after self-calibration can often differ from their a priori values by $\gg$10\% in EHT data \citep[see, e.g.,][]{M87Paper4}, which substantially exceeds the differences between the reported SEFDs and those predicted by \ngehtsim.

\subsection{EHT observations at 0.87\,mm} \label{sec:EHT345}

Many of the EHT stations are equipped with receivers capable of observing at a wavelength of 0.87\,mm, and the EHT has already exercised this capability during test observations in October 2018 \citep{Raymond_2024} and April 2021 (observing \m87) and as part of a science campaign in April 2023 (observing \sgra).  As of the time of writing for this paper, none of the data from the 2021 or 2023 observations have yet been published or publicly released, but we can nevertheless use \ngehtsim to simulate the expected data quality of these and future 0.87\,mm observations with the EHT.  We describe and present such simulations in this section.

\begin{deluxetable*}{lccc}
\tablecolumns{4}
\tablewidth{0pt}
\tablecaption{Simulation parameters for EHT 0.87\,mm observations\label{tab:EHT345GHz}}
\tablehead{\colhead{Parameter} & \colhead{EHT 2021} & \colhead{EHT 2023} & \colhead{EHT near-future}}
\startdata
\multirow{2}{*}{sites} & ALMA, GLT, JCMT, & ALMA, APEX, & ALMA, APEX, GLT, JCMT, LMT, \\
 & NOEMA, PV, SMA, SMT & JCMT, SMA, SMT & NOEMA, PV, SMA, SMT, SPT \\
target source & \m87 & \sgra & \m87, \sgra \\
frequency & 337.6\,GHz & 337.6\,GHz & 337.6\,GHz \\
bandwidth & 2\,GHz & 2\,GHz & 2\,GHz \\
date & April 19, 2021 & April 15, 2023 & April \\
starting time & 1\,UT & 2\,UT & 1\,UT \\
track duration & 5 hours & 13 hours & 14 hours \\
\enddata
\tablecomments{Parameters determining the structure of the observing tracks for our simulations of EHT observations at 0.87\,mm observing wavelength.}
\end{deluxetable*}

\begin{figure*}
    \centering
    \includegraphics[width=1.00\textwidth]{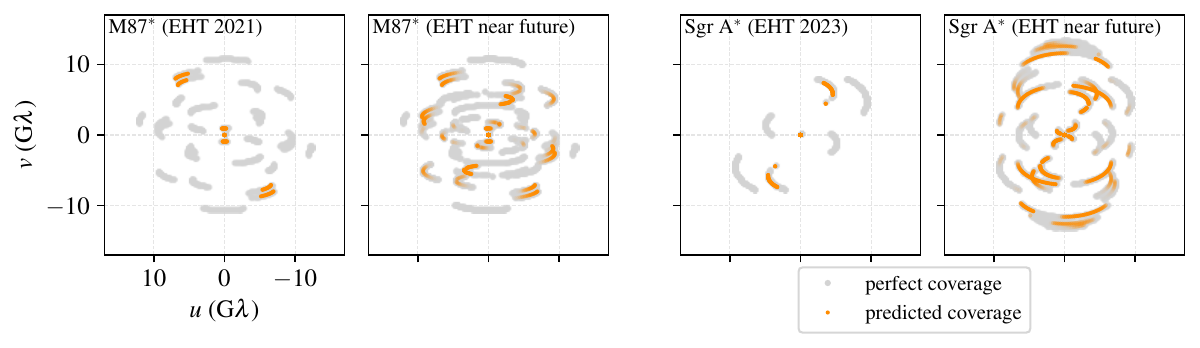}
    \caption{Simulated \uv-coverage for past and near-future EHT observations at an observing wavelength of 0.87\,mm; see \autoref{sec:EHT345} for simulation details.  The leftmost panel shows the coverage predicted for an EHT observation of \m87 carried out on April 19, 2021, and the second panel shows the coverage predicted for a near-future version of the EHT in which the LMT and APEX telescopes are also able to join the \m87 observation at 0.87\,mm.  The third panel shows the coverage predicted for an EHT observation of \sgra carried out on April 15, 2023, and the rightmost panel shows the coverage predicted for a near-future version of the EHT in which the LMT, NOEMA, PV, and SPT telescopes are also able to join the \sgra observation at 0.87\,mm.  In all panels, gray points indicate visibilities that would have been detected if the array were observing with infinite sensitivity, and orange points indicate visibilities that \ngehtsim predicts should be detected given the finite sensitivity of the real array.  For the first and third panels from the left, we assume observing conditions appropriate for the specific dates April 19, 2021 and April 15, 2023, respectively.  For the second and fourth panels from the left, we simulate 100 different instantiations of April weather conditions, and the opacity of each plotted point is proportional to how frequently it is detected.  All panels share the same horizontal and vertical axis ranges, which are explicitly labeled in the left panel.} \label{fig:coverage_EHT_345GHz}
\end{figure*}

\subsubsection{2021 EHT observations of \m87} \label{sec:EHT2021}

To simulate the 2021 EHT observations of \m87 at 0.87\,mm, we use the settings specified in the second column of \autoref{tab:EHT345GHz}.  For the source structure, we use the same geometric model as in \autoref{sec:EHT2017}, which is detailed in \autoref{app:SourceModels}, and the assumed receiver and dish properties for each participating telescope are detailed in \autoref{app:Receivers} and \autoref{app:SiteTable}, respectively.  As in \autoref{sec:EHT2017}, we simulate atmospheric conditions at each site that are specific to the April 19, 2021 observing date.

The left panel of \autoref{fig:coverage_EHT_345GHz} shows the \uv-coverage predicted by \ngehtsim (in orange) compared with the coverage that would be obtained by the same array observing with infinite sensitivity (in gray).  The simulation predicts that of the 16 distinct baselines that could have been detected by an arbitrarily sensitive version of the array, only four baselines -- ALMA-PV, ALMA-NOEMA, NOEMA-PV, and the ``zero baseline'' JCMT-SMA -- are actually able to recover detections under the specified observing conditions.

\subsubsection{2023 EHT observations of \sgra} \label{sec:EHT2023}

To simulate the 2023 EHT observations of \sgra at 0.87\,mm, we use the settings specified in the third column of \autoref{tab:EHT345GHz}.  For the source structure, we use a similar geometric model as in \autoref{sec:EHT2017} and \autoref{sec:EHT2021}, but with parameters that have been chosen to mimic the observed horizon-scale \sgra structure and with the addition of interstellar scattering effects; \autoref{app:SourceModels} provides a detailed description of the source model.  As in the previous sections, we use the receiver and dish properties detailed in \autoref{app:Receivers} and \autoref{app:SiteTable}, respectively, and we simulate atmospheric conditions at each site that are specific to the April 15, 2023 observing date.

The third panel from the left panel in \autoref{fig:coverage_EHT_345GHz} shows the \uv-coverage predicted by \ngehtsim (in orange) compared with the coverage that would be obtained by the same array observing with infinite sensitivity (in gray).  Though five stations took part in this observation, two pairs of them -- ALMA-APEX (in Chile) and JCMT-SMA (in Hawaii) -- are co-located and thus do not contribute geometrically unique baseline coverage.  I.e., the array was effectively operating as a three-station one, plus the addition of a ``zero baseline'' comprised of the co-located sites.  The simulation predicts that of the four distinct baselines that could have been detected by an arbitrarily sensitive version of the array, only two of them -- Chile-SMT and the zero baseline -- are actually able to recover detections under the specified observing conditions.

\subsubsection{Near-future EHT observations at 0.87\,mm} \label{sec:NearFutureEHT}

Only a subset of the EHT array was able to participate in the 2021 and 2023 observations at 0.87\,mm, limiting the \uv-coverage that could be achieved.  In the near future, the EHT may carry out observations at 0.87\,mm that feature a more complete array, including sites that do not currently have 0.87\,mm receivers but which are planning to acquire them.  We can use \ngehtsim to simulate the expected performance of the array during such observations of \m87 and \sgra.  Our simulations use the settings specified in the fourth column of \autoref{tab:EHT345GHz}.

As in previous sections, we use the receiver and dish properties detailed in \autoref{app:Receivers} and \autoref{app:SiteTable}, respectively.  For the LMT and SPT telescopes -- which are not currently equipped with 0.87\,mm receivers -- we assume specifications that match those of ALMA (i.e., a receiver temperature of $T_{\text{rx}} = 75$\,K and a sideband separation ratio of $r = 0.1$).  \autoref{app:SourceModels} provides a detailed description of the source models for both \m87 and \sgra.  Because we do not know what the exact weather conditions will be for future observations, we instead simulate atmospheric conditions at each site by randomly selecting a past date and using the conditions from that date.  For each observation, we run 100 simulations using different such samples of the historical global weather conditions in April, which permits us to evaluate the typical expected performance of the array.

The second and fourth panels from the left in \autoref{fig:coverage_EHT_345GHz} show the \uv-coverage predicted by \ngehtsim (in orange) compared against the coverage that would be obtained by the same array observing with infinite sensitivity (in gray) for the \m87 and \sgra simulations, respectively.  The 100 weather instantiations are represented using the opacity of the plotted orange points, such that visibilities that are detected 100\% of the time are fully opaque and visibilities that are detected 0\% of the time are fully transparent (visibilities that are detected some fraction of the time are plotted with the corresponding fractional opacity).  For both the \m87 and \sgra observations, we see that the increased site participation noticeably improves the \uv-coverage.  However, it remains the case -- particularly for \m87 -- that a substantial fraction of the baselines do not achieve detections.

For \m87, only a fraction $0.23 \pm 0.08$ of all baselines are detected, where we report the median and one standard deviation from the 100 weather instantiations; when considering only the geometrically unique baselines (i.e., consolidating redundant baselines such as ALMA-SMT and APEX-SMT), the detection fraction is $0.17 \pm 0.07$.  The corresponding fractions for \sgra are $0.40 \pm 0.08$ for all baselines and $0.36 \pm 0.11$ for the geometrically unique baselines.  Compared to analogous EHT observations at 1.3\,mm observing wavelength -- which are predicted to achieve typical detection fractions of ${>}0.97$ for both sources and both baseline groupings -- the 0.87\,mm observations suffer from a high fractional loss of detections.

\subsection{Multi-frequency observations with a next-generation EHT} \label{sec:EHTmultifreq}

As shown in the previous section, 0.87\,mm observations carried out by the current and near-future EHT array are expected to achieve detection rates of no more than $\sim$50\%.  A key limitation driving this low detection fraction is the requirement for baselines to achieve sufficient sensitivity to track atmospheric phase variations within a fraction of the short ($\sim$20-second) coherence timescale at 0.87\,mm.  A promising avenue towards realizing longer integration times is the frequency phase transfer (FPT) technique, whereby atmospheric phases tracked at some ``reference'' frequency -- typically a lower frequency, where the dimensionless baseline lengths are shorter and the array more sensitive -- can be transferred to simultaneous observations made at a ``target'' frequency \citep[see, e.g.,][]{Rioja_2020}.  As described in \autoref{sec:VisDetection} (see also \autoref{sec:PhaseTrackingMultiFreq}), \ngehtsim can simulate observations carried out using the FPT technique.  In this section, we demonstrate the impact that FPT could have on 0.87\,mm EHT observations of \m87 and \sgra.

\begin{deluxetable*}{lcc}
\tablecolumns{3}
\tablewidth{0pt}
\tablecaption{Simulation parameters for multi-frequency EHT observations\label{tab:FPT_345GHz}}
\tablehead{\colhead{Parameter} & \colhead{\m87 observations} & \colhead{\sgra observations}}
\startdata
\multirow{2}{*}{sites} & ALMA, APEX, GLT, JCMT, & ALMA, APEX, JCMT, LMT, \\
 & LMT, NOEMA, PV, SMA, SMT & NOEMA, PV, SMA, SMT, SPT \\
target source & \m87 & \sgra \\
frequency & 86\,GHz, 345\,GHz & 230\,GHz, 345\,GHz \\
bandwidth & 2\,GHz & 2\,GHz \\
date & April & April \\
starting time & 1\,UT & 1\,UT \\
track duration & 14 hours & 14 hours \\
dual-band sites & APEX, GLT, JCMT, LMT, SMT & APEX, JCMT, LMT, SMT, SPT \\
single-band sites & ALMA, NOEMA, PV, SMA & ALMA, NOEMA, PV, SMA \\
\enddata
\tablecomments{Parameters determining the structure of the observing tracks for our simulations of a future version of the EHT that observes at 0.87\,mm using FPT from either 3\,mm or 1.3\,mm.  The listed dual-band sites are assumed to observe at both specified frequencies, while the single-band sites are assumed to observe only at the higher frequency.}
\end{deluxetable*}

A number of EHT sites are planning to implement tri-band observing capabilities as part of a next-generation EHT upgrade \citep{Doeleman_2023}, with a complement of receivers covering the 86\,GHz (3\,mm), 230\,GHz (1.3\,mm), and 345\,GHz (0.87\,mm) bands.  Our simulations in this section adhere to the expected multi-frequency capabilities following these upgrades; for the simulations of \m87 and \sgra we use the settings specified in the second and third columns of \autoref{tab:FPT_345GHz}, respectively.  For the \m87 simulations, we assume that FPT will be carried out using the 3\,mm band as the reference frequency, while for \sgra we assume that the 1.3\,mm band will be the reference frequency (to avoid the substantial effects of interstellar scattering at 3\,mm).  We continue to use the dish specifications provided in \autoref{app:SiteTable}, and we assume the 1.3 and 0.87\,mm receiver specifications detailed in \autoref{app:Receivers} (except for the LMT and SPT telescopes at 0.87\,mm, for which we assume specifications that match those of ALMA, as in \autoref{sec:NearFutureEHT}).  For all sites observing with a 3\,mm receiver band, we assume ALMA-like specifications, corresponding to a receiver temperature of $T_{\text{rx}} = 40$\,K and a sideband separation ratio of $r = 0.03$ \citep{Claude_2008}.  Note that although the upgrades specified in \citet{Doeleman_2023} include increased bandwidths alongside the multi-frequency capabilities, we continue to use 2\,GHz bandwidths for the simulations in this section so as to isolate the impact of FPT and enable more direct comparisons with the results from prior sections (for analogous simulations that use wider bandwidths, see \autoref{sec:HighFreqM87SgrA} and \autoref{sec:HighFreqPS}).  We also continue to use the source models described in \autoref{app:SourceModels} for both \m87 and \sgra.

\begin{figure}
    \centering
    \includegraphics[width=1.00\columnwidth]{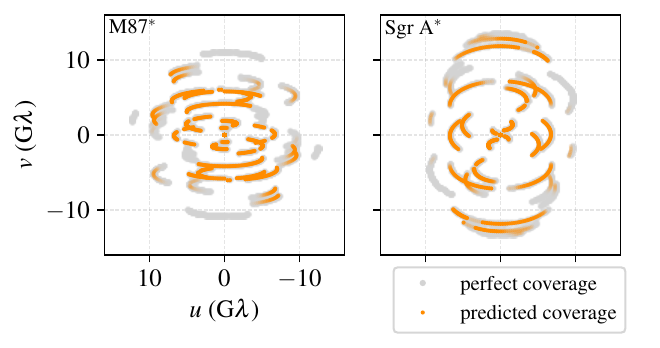}
    \caption{Simulated \uv-coverage for a future version of the EHT observing at a wavelength of 0.87\,mm and utilizing the FPT technique; see \autoref{sec:EHTmultifreq} for simulation details.  The expected typical coverage achieved for observations of \m87 is shown on the left, and the corresponding coverage achieved for observations of \sgra is shown on the right.  The plotting style is analogous to that in \autoref{fig:coverage_EHT_345GHz}, with the opacity of each orange point corresponding to how frequently the corresponding visibility is detected.} \label{fig:coverage_FPT_345GHz}
\end{figure}

As in \autoref{sec:NearFutureEHT}, we run 100 simulations for both \m87 and \sgra, using different samples of the historical global weather conditions in April.  \autoref{fig:coverage_FPT_345GHz} shows the resulting \uv-coverage predicted by \ngehtsim (in orange) compared against the coverage that would be obtained by the same array observing with infinite sensitivity (in gray).  For both the \m87 and \sgra observations, we see that the addition of FPT noticeably improves the \uv-coverage relative to the corresponding panels of \autoref{fig:coverage_EHT_345GHz}.  For \m87, a fraction $0.55 \pm 0.06$ of all baselines are detected, and a fraction $0.58 \pm 0.06$ of geometrically unique baselines are detected.  The corresponding fractions for \sgra are $0.55 \pm 0.06$ for all baselines and $0.54 \pm 0.08$ for the geometrically unique baselines.

\subsection{VLBI at very high frequencies} \label{sec:HighFrequencyVLBI}

Though VLBI has not yet been carried out at wavelengths shorter than 0.87\,mm, there are a number of atmospheric windows -- such as those around 0.65\,mm (460\,GHz), 0.43\,mm (690\,GHz), and 0.34\,mm (875\,GHz) -- that are accessible to high-elevation facilities (e.g., ALMA) and which could seemingly permit VLBI observations.  Not many (sub)millimeter facilities are currently equipped with appropriate receivers for observing at these higher frequencies, but it would in principle be possible to outfit them appropriately, and so we can nevertheless use \ngehtsim to explore the prospects for VLBI at these frequencies.

\subsubsection{Single-baseline sensitivity} \label{sec:HighFreqSingleBaseline}

\begin{figure*}
    \centering
    \includegraphics[width=1.00\textwidth]{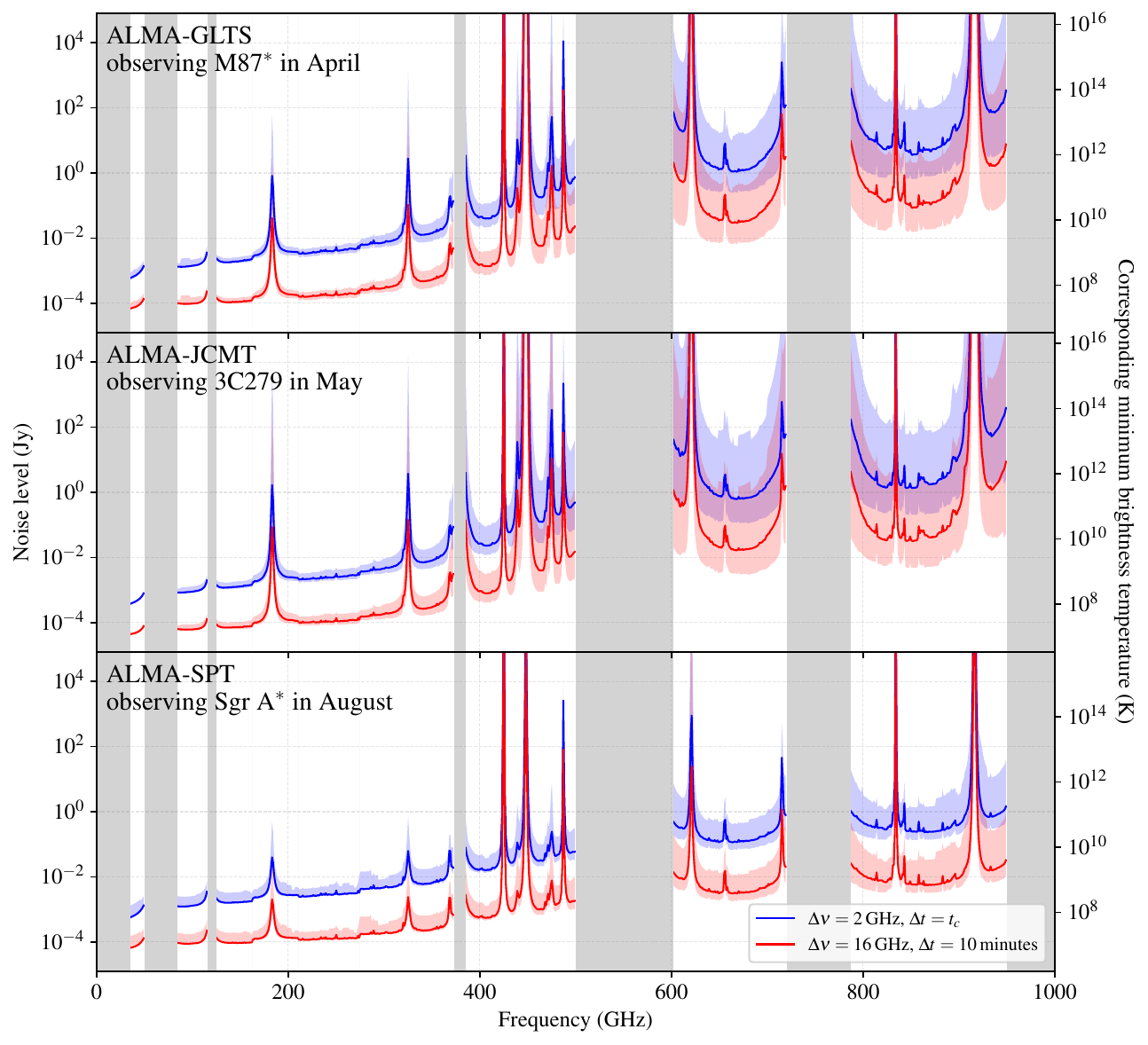}
    \caption{Thermal noise level achievable on three different baselines as a function of observing frequency; see \autoref{sec:HighFreqSingleBaseline} for simulation details.  Each of these baselines has been selected because it connects two sites with exceptionally dry observing conditions, and we have selected the times of year and local observing times to optimize the quality of the weather for each of these simulations.  Blue curves indicate performance expectations for capabilities that are comparable to those currently used by the EHT, while red curves indicate performance expectations for capabilities that substantially exceed those of current arrays but which are similar to the planned capabilities of a next-generation EHT \citep{Doeleman_2023}.  In all cases, the central line shows the median RMS thermal noise achieved from 100 instantiations of weather conditions appropriate for the specified month, and the lighter shaded region brackets the 90th percentile interval (i.e., spanning the 5th to the 95th percentile values).  The darker gray vertical shaded regions indicate spectral ranges not currently covered by ALMA receiver bands.  Large spikes in the RMS noise level correspond to strong atmospheric lines (see, e.g., \autoref{fig:resolution_comparison}).} \label{fig:noise_spectrum}
\end{figure*}

\autoref{fig:noise_spectrum} shows the thermal noise level achievable on three specific baselines -- ALMA-GLTS\footnote{The GLTS site is located at the summit of the Greenland ice sheet, with a latitude of $72.580^{\circ}$, a longitude of $-38.449^{\circ}$, and an elevation of 3230\,meters. The Greenland telescope (GLT) plans to relocate to the summit site, which has substantially drier atmospheric conditions than the current site \citep{Matsushita_2017}.  For these simulations, we thus retain the dish specifications provided in \autoref{app:SiteTable} but use the GLTS rather than the GLT site.}, ALMA-JCMT, and ALMA-SPT -- as a function of observing frequency.  Each of these baselines connects two stations that can experience exceptionally dry atmospheric conditions, and for each baseline we have simulated the noise levels during the time of year expected to minimize the baseline thermal noise.  We have also fixed $\eta_w = 1$ (see \autoref{eqn:SEFD}), equivalent to assuming negligible wind at all sites.  These baselines are thus meant to represent the ``best case'' for high-frequency VLBI detection prospects using current or near-future facilities.

We simulate the ALMA-GLTS baseline observing \m87 at 3\,UT in April, corresponding to a source elevation of $\sim$54\,degrees at ALMA and $\sim$28.5\,degrees at GLTS.  We simulate the ALMA-JCMT baseline observing 3C279 at 5\,UT in May, corresponding to a source elevation of $\sim$42\,degrees at ALMA and $\sim$42\,degrees at JCMT.  We simulate the ALMA-SPT baseline observing \sgra at 0\,UT in August, corresponding to a source elevation of $\sim$79\,degrees at ALMA and $\sim$29\,degrees at SPT.  For each of these simulations, we have assumed that both participating stations are equipped with an ALMA-like receiver suite whose specifications are listed in \autoref{tab:HighFrequency}.  The sensitivity curves in \autoref{fig:noise_spectrum} are shown for two different choices of bandwidth and integration time: (1) a bandwidth of $\Delta \nu = 2$\,GHz and an integration time of $\Delta t = t_c$ (i.e., similar to the values that are most appropriate for current EHT observations) is shown in blue, and (2) a bandwidth of $\Delta \nu = 16$\,GHz and an integration time of $\Delta t = 10$\,minutes (i.e., similar to the values that might be relevant for a next-generation EHT employing the FPT technique) is shown in red.  As in \autoref{sec:NearFutureEHT} and \autoref{sec:EHTmultifreq}, we sample 100 instantiations of the atmospheric conditions for each simulation, using different samples of the historical global weather conditions for the appropriate month; the corresponding range of baseline performance is indicated by the shaded region around each sensitivity curve in \autoref{fig:noise_spectrum}.

\begin{deluxetable}{lcccc}
\tablecolumns{5}
\tablewidth{0pt}
\tablecaption{ALMA receiver properties\label{tab:HighFrequency}}
\tablehead{\colhead{Band} & \colhead{Frequency} & \colhead{$T_{\text{rx}}$} & \colhead{$r$} & \colhead{References}}
\startdata
Band 1 & 35--50\,GHz & 25\,K & 0.1 & \citet{Huang_2018} \\
Band 3 & 84--116\,GHz & 40\,K & 0.03 & \citet{Claude_2008} \\
Band 4 & 125--163\,GHz & 40\,K & 0.1 & \citet{Asayama_2008} \\
Band 5 & 163--211\,GHz & 55\,K & 0.1 & \citet{Billade_2012} \\
Band 6 & 211--275\,GHz & 40\,K & 0.01 & \citet{Kerr_2004} \\
Band 7 & 275--373\,GHz & 75\,K & 0.1 & \citet{Mahieu_2012} \\
Band 8 & 385--500\,GHz & 150\,K & 0.1 & \citet{Sekimoto_2008} \\
Band 9 & 602--720\,GHz & 100\,K & 1 & \citet{Baryshev_2008} \\
Band 10 & 787--950\,GHz & 100\,K & 1 & \citet{Fujii_2013} \\
\enddata
\tablecomments{Receiver properties for the ALMA receiver suite, which are assumed for the simulations carried out in \autoref{sec:HighFrequencyVLBI}.  $T_{\text{rx}}$ is the receiver temperature and $r$ is the sideband separation ratio (see \autoref{eqn:SystemTemperature}).}
\end{deluxetable}

For single-baseline measurements and assuming a circularly-symmetric Gaussian source structure, we can associate a minimum detectable brightness temperature with the thermal noise limit using \citep[see][]{Lobanov_2015}

\begin{align}
T_{b,\text{min}} & = \frac{\pi e \sigma_b b^2}{2 k} \nonumber \\
& \approx \big( 3.1 \times 10^{11} \text{ K} \big) \left( \frac{\sigma_b}{1 \text{ Jy}} \right) \left( \frac{b}{10^4 \text{ km}} \right)^2 , \label{eqn:MinTb}
\end{align}

\noindent where $\sigma_b$ is the thermal noise level on the baseline, $b$ is the projected baseline length, $k$ is the Boltzmann constant, and $e$ is the mathematical constant corresponding to the base of the natural logarithm.  The minimum brightness temperature is indicated by the right-hand vertical axis for each of the panels in \autoref{fig:noise_spectrum}.

It is evident from \autoref{fig:noise_spectrum} that the noise level rises rapidly towards higher observing frequencies.  At the frequencies and capabilities of interest for current and near-future EHT observations -- i.e., observing frequencies up to $\sim$345\,GHz, bandwidths of $\sim$2\,GHz and integration times limited by the coherence time -- the baseline thermal noise levels are typically below $\sim$20\,mJy.  At higher frequencies (e.g., corresponding to ALMA Band 9 and Band 10) and assuming the same capabilities, baseline thermal noise levels never get better than $\sim$100\,mJy, with typical values being $\sim$1--10\,Jy for ALMA-GLTS and ALMA-JCMT and $\sim$150--300\,mJy for ALMA-SPT.  When assuming capabilities that are more appropriate for a next-generation EHT -- i.e., bandwidths of $\sim$16\,GHz and integration times of $\sim$10\,minutes -- the expected baseline sensitivity improves by $\sim$1--2 orders of magnitude, reaching as low as $\sim$10\,mJy for ALMA-GLTS and ALMA-JCMT and $\sim$2--3\,mJy for ALMA-SPT in the highest observing bands.

\subsubsection{Full-array VLBI of \m87 and \sgra} \label{sec:HighFreqM87SgrA}

\begin{figure*}
    \centering
    \includegraphics[width=0.90\textwidth]{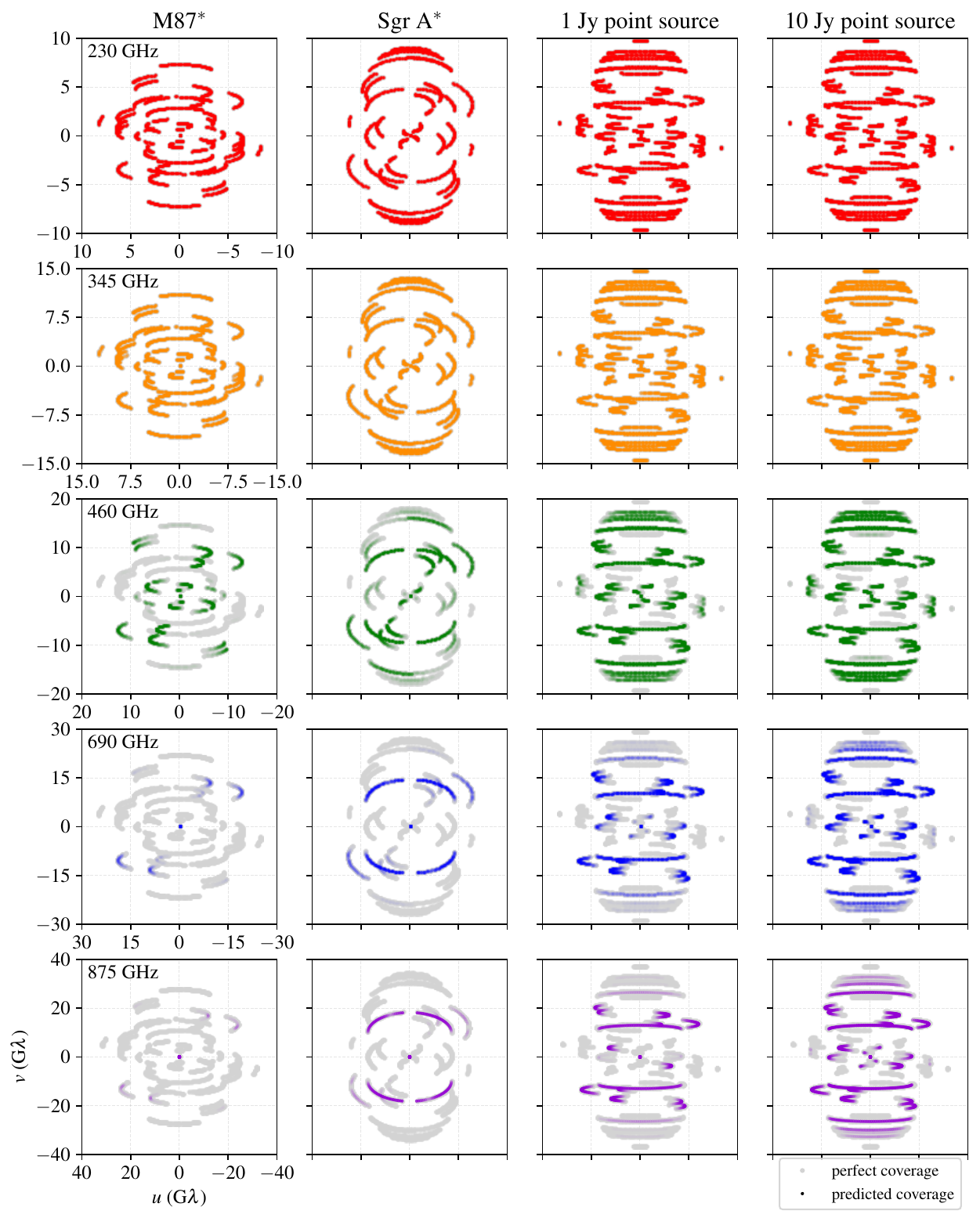}
    \caption{Simulated \uv-coverage plots for a hypothetical future EHT array capable of observing at multiple different frequencies (see \autoref{sec:HighFrequencyVLBI}).  From left to right, the columns correspond to observations of \m87, \sgra, a 1\,Jy point source, and a 10\,Jy point source; the point sources are assumed to be located at the position 3C279.  From top to bottom, the rows correspond to observing frequencies of 230\,GHz, 345\,GHz, 460\,GHz, 690\,GHz, and 875\,GHz.  The plotting style is analogous to that in \autoref{fig:coverage_EHT_345GHz}, with the opacity of each colored point corresponding to how frequently the corresponding visibility is detected.  Each row of panels shares common horizontal and vertical axis ranges, which are explicitly labeled in the leftmost panel of the row.
    } \label{fig:coverage_high-frequency}
\end{figure*}

The example baseline noise levels described in the previous section indicate that long-baseline observations at high observing frequencies could in principle achieve astrophysically relevant sensitivities, given the wide bandwidths and long integration times that are expected to be accessible to a next-generation EHT (and assuming that the sites are appropriately outfitted with the necessary receivers).  However, the simulations in \autoref{sec:HighFreqSingleBaseline} are carried out on a per-baseline level, with the time of year and source location in the sky selected to optimize the performance of each baseline individually.  For observations carried out using a more complete array, such optimization is not possible.  Furthermore, the simulations in \autoref{sec:HighFreqSingleBaseline} only compute the achievable baseline noise level, and they do not take into account the corresponding expected source flux density on each baseline (which is necessary to determine detectability, per \autoref{sec:VisDetection}).  In this section, we carry out high-frequency VLBI simulations of a futuristic version of the EHT array that is fully equipped with an ALMA-like suite of receivers.

For the simulations in this section, we assume that the entire EHT array has been equipped with a receiver suite whose specifications are listed in \autoref{tab:HighFrequency}.  We simulate April observations of both \m87 and \sgra at observing frequencies of 230, 345, 460, 690, and 875\,GHz, assuming a bandwidth of 16\,GHz and an integration time of 10 minutes for all simulations.  As in \autoref{sec:EHTmultifreq}, we assume that FPT will be carried out using the 3\,mm band as the reference frequency for \m87 simulations, while for \sgra we assume that the 1.3\,mm band will be the reference frequency.  We continue to use the dish specifications provided in \autoref{app:SiteTable} and the source models described in \autoref{app:SourceModels} for both \m87 and \sgra, with the amount of interstellar scattering applied to the \sgra model adjusted appropriately for each simulated observing frequency.  As in previous sections, we sample 100 instantiations of the atmospheric conditions for each simulation, using different samples of the historical global weather conditions for April.

\autoref{fig:coverage_high-frequency} shows the \uv-coverage predicted by \ngehtsim (in color) compared against the coverage that would be obtained if the EHT were observing with infinite sensitivity (in gray) at all five observing frequencies.  The \m87 simulations are shown in the left column and the \sgra simulations are shown in the second column from the left.  For the colored visibilities -- i.e., those labeled as an ``achieved detection'' in the figure -- we plot only those detections that achieve an SNR of ${\geq} 1$ within the 10-minute integration time.  For both \m87 and \sgra observations, we see that the 230 and 345\,GHz observations consistently achieve a nearly $\sim$100\% detection rate, while the detection fraction drops off considerably at the higher frequencies.  The 460\,GHz simulations typically achieve detection fractions on non-intrasite baselines of $0.23 \pm 0.09$ and $0.29 \pm 0.10$ for \m87 and \sgra, respectively, with the SNR on some of these baselines regularly exceeding a value of 10.  For 690 and 875\,GHz observations, it is common to have only one or zero non-intrasite baselines detected.  The SNR on non-intrasite baselines for \m87 observations at 690 and 875\,GHz does not exceed 3 and 2, respectively.  For \sgra observations at 690 and 875\,GHz, the ALMA-SPT baseline is by far the most frequently detected, achieving peak SNR values of $\sim$9 and $\sim$6, respectively.

\subsubsection{Full-array VLBI of the brightest sources} \label{sec:HighFreqPS}

In addition to the substantially increased atmospheric optical depths at higher observing frequencies, another limiting factor is the modest correlated flux densities of \m87 and \sgra on the longest baselines.  In our models, \m87 and \sgra have long-baseline ($\sim$30\,G$\lambda$) flux densities of $\sim$10\,mJy and $\sim$20\,mJy, respectively.  Other sources observed with the EHT -- including 3C279 \citep{Kim_2020}, Centaurus A \citep{Janssen_2021}, J1924-2914 \citep{Issaoun_2022}, and NRAO 530 \citep{Jorstad_2023} -- exhibit flux densities between $\sim$100\,mJy and $\sim$500\,mJy on $\sim$10\,G$\lambda$ baselines.  Furthermore, the brightness temperature maximum expected for synchrotron emission is ${\sim}10^{12}$\,K \citep{Kellermann_1969}, corresponding (per \autoref{eqn:MinTb}) to correlated flux densities between $\sim$1\,Jy and $\sim$10\,Jy on the longest baselines.\footnote{We note that there are fewer than 200 compact sources in the \textit{Planck} 857\,GHz catalog with flux densities above 1\,Jy \citep{Planck_2016}, where ``compact'' for \textit{Planck} means smaller than its roughly 4.5-arcminute beam.  Of these sources, none have spectral indices ($S_{\nu} \propto \nu^{\alpha}$) smaller than $\alpha \approx 0.8$, indicating that the population is likely to be dominated by dusty galaxies rather than synchrotron sources.}  The detection prospects for VLBI observations of such sources could thus be substantially better than for \m87 or \sgra.

Using the same array setup as in \autoref{sec:HighFreqM87SgrA}, we simulate observations of a hypothetical object that has a point-source emission structure and which is located at the near-equatorial position of the radio source 3C279.  The right two columns of \autoref{fig:coverage_high-frequency} show the resulting \uv-coverages for these observations assuming (second column from the right) a flux density of 1\,Jy and (rightmost column) a flux density of 10\,Jy.  We see that even for the maximally optimistic case of a 10\,Jy point source, the detection fraction begins to drop noticeably for observing frequencies above 345\,GHz.  At the highest simulated observing frequency of 875\,GHz, the detection fraction on non-intrasite baselines is $0.13 \pm 0.06$ and $0.27 \pm 0.08$ for the 1\,Jy and 10\,Jy point source, respectively.

\section{Summary and conclusions} \label{sec:Summary}

In this paper we present \ngehtsim, a Python-based software package for generating realistic synthetic data appropriate for high-frequency VLBI observations.  \ngehtsim includes a database of historical atmospheric information, which is tabulated for several dozen existing radio and (sub)millimeter telescope sites using more than a decade of MERRA-2 atmospheric state data processed through the \am radiative transfer code.  Synthetic observations generated with \ngehtsim combine the resulting atmospheric optical depth and brightness temperature information with telescope and receiver specifications to determine baseline sensitivities across the array, from which a heuristic algorithm that emulates both single- and multi-frequency fringe-finding techniques determines whether individual visibilities are detected.

We demonstrate the capabilities of \ngehtsim by generating a series of example synthetic EHT observations of \m87 and \sgra, two sources for which we have approximate knowledge of their structure on $\sim$microarcsecond scales.  Synthetic observations generated by \ngehtsim are able to reproduce well the 2017 EHT data that yielded the first published images of the \m87 and \sgra black holes at an observing frequency of 230\,GHz.  Using \ngehtsim to simulate past and future EHT observations at 345\,GHz, we show that the 2021 and 2023 EHT observations of \m87 and \sgra, respectively, will likely have low detection fractions and correspondingly poor \uv-coverage (see first and third panels of \autoref{fig:coverage_EHT_345GHz}).  Single-frequency image reconstructions will not be possible using these datasets.  Near-future EHT observations of both \m87 and \sgra at 345\,GHz have the possibility to perform considerably better than prior observations, but they will continue to suffer from $<$50\% detection fractions and much poorer \uv-coverage than comparable observations at 230\,GHz (see second and fourth panels of \autoref{fig:coverage_EHT_345GHz}).

The addition of simultaneous multi-frequency observing capabilities to the EHT array can improve its detection prospects at 345\,GHz through the use of the FPT calibration technique, yielding detection fractions that are consistently $>$50\% (see \autoref{fig:coverage_FPT_345GHz}).  Further improving the baseline sensitivity through, e.g., bandwidth upgrades across the array can bring the detection fraction at 345\,GHz up to nearly 100\%, matching the performance at 230\,GHz (see \autoref{fig:coverage_high-frequency}).

Anticipating the continuation of historical trends to push VLBI towards ever-higher observing frequencies, we use \ngehtsim to simulate futuristic EHT-like observations at frequencies above 345\,GHz.  Given sufficiently wide bandwidths ($\sim$16\,GHz) and long integration times ($\sim$10\,minutes), baselines connecting exceptionally dry sites -- such as ALMA, GLTS, JCMT, and SPT -- can achieve astrophysically relevant sensitivities at observing frequencies that fall in atmospheric windows around, e.g., 460, 690, and 875\,GHz (see \autoref{fig:noise_spectrum}).  However, array-wide observations of sources such as \m87 and \sgra exhibit heavily degraded performance at these higher frequencies.  Observations of \m87 and \sgra with a futuristic EHT array that is appropriately outfitted to observe at 460\,GHz could expect to regularly achieve multiple detections on long baselines, though with a detection fraction that does not exceed $\sim$30\% (see \autoref{fig:coverage_high-frequency}).  Analogous observations at 690 and 875\,GHz consistently see almost no detections at all beyond those on the single baseline ALMA-SPT, and even that baseline does not achieve signal-to-noise ratios above 10 in 10-minute integration times.  High-frequency observations of continuum sources that are substantially brighter than either \m87 or \sgra on long baselines (i.e., correlated flux densities $\gtrsim$1\,Jy) are viable in principle, though detection fractions remain low ($\lesssim$30\%) and it is unclear whether a population of sufficiently bright sources actually exists.

We note that one of the most important assumptions underpinning the simulations carried out in this paper is the value of the atmospheric coherence time at each observing frequency.  For the simulations presented here we have assumed a fixed set of coherence times, which have been selected to be characteristic of atmospheric conditions at millimeter-wavelength sites such as those participating in EHT observations.  However, the specific values of the coherence times at each site are not known, and the quantitative details of the simulation predictions (e.g., detection fractions) depend -- sometimes sensitively -- on the assumed coherence times.  The FPT calibration technique can mitigate the impact of an uncertain coherence time at the target frequency, but it still relies on knowledge of the coherence time at the reference frequency.  While \ngehtsim permits manual exploration of different coherence time assumptions, it does not automatically adjust the coherence time based on local weather conditions; future work is necessary to understand whether -- and if so, how -- local atmospheric state or other accessible physical information may be used to produce reasonable estimates of the coherence time.

We close by noting that although \ngehtsim has been developed primarily for predicting current and next-generation EHT performance, it is a general-purpose tool that can readily simulate the performance of other existing or future arrays.  As such, \ngehtsim can be used for applications such as determining antenna placement during the design of future arrays, predicting array performance for observing proposals submitted to existing arrays, or supplying weather-informed telescope sensitivity estimates for initial flux density calibration of collected data.

\software{\am \citep{Paine_2022}, \texttt{ehtim} \citep{Chael_2016,Chael_2018,Chael_2023}, \texttt{ehtplot} \citep{Chan_2021}, \texttt{matplotlib} \citep{Hunter_2007}, \texttt{netCDF4} \citep{Whitaker_2020}, \ngehtsim \citep{Pesce_2024}, \texttt{numpy} \citep{Harris_2020}, \texttt{paramsurvey}\footnote{\url{https://github.com/wumpus/paramsurvey}}, \texttt{ray} \citep{Moritz_2017}, stochastic optics \citep{Johnson_2016}, \texttt{scipy} \citep{Virtanen_2020}}

\acknowledgments

We thank Avery Broderick, Dominic Chang, Roger Deane, Richard Dodson, Garret Fitzpatrick, Boris Georgiev, Linus Hamilton, Johnson Han, Aaron Oppenheimer, Nimesh Patel, Alex Raymond, Mar\'ia Rioja, Senkhosi Simelane, Ranjani Srinivasan, and Maciek Wielgus for various conversations and discussions that have been helpful throughout the course of this project.  We also thank an anonymous referee for providing constructive comments that improved the quality of the paper.

Topographical information used in this paper has been obtained from the TessaDEM database\footnote{\url{https://tessadem.com/}}, which is licensed under the Open Database License (ODbL) v1.0.  This research has made use of the NASA/IPAC Extragalactic Database (NED), which is funded by the National Aeronautics and Space Administration and operated by the California Institute of Technology.  This research has made use of the SIMBAD database,
operated at CDS, Strasbourg, France.

Support for this work was provided by the NSF (AST-1935980, AST-2034306, and AST-2307887) and by the Gordon and Betty Moore Foundation through grant GBMF-10423.  This work has been supported in part by the Black Hole Initiative at Harvard University, which is funded by grants from the John Templeton Foundation (Grant \#62286) and the Gordon and Betty Moore Foundation (Grant GBMF-8273).

\clearpage

\bibliography{references}{}
\bibliographystyle{aasjournal}

\appendix
\numberwithin{equation}{section}

\section{Spectral resolution} \label{app:Resolution}

\begin{figure*}
    \centering
    \includegraphics[width=1.00\textwidth]{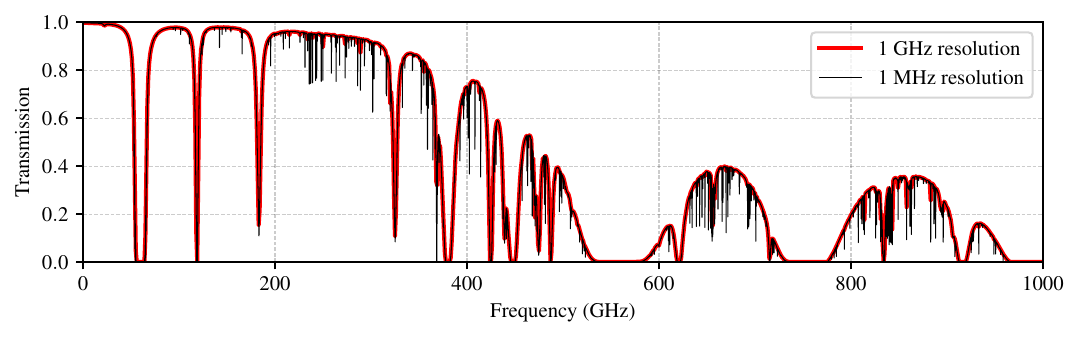}
    \caption{Comparison of the zenith atmospheric transmission spectrum over the ALMA site, computed using \am set with two different spectral resolutions.  The red spectrum is computed with a 1\,GHz spacing between spectral points, and the black spectrum is computed with a 1\,MHz spacing between spectral points. The atmospheric layer settings used to specify the \am input for these spectra correspond to the April 15, 2022 MERRA-2 reanalysis at 0 UTC.}\label{fig:resolution_comparison}
\end{figure*}

\autoref{fig:resolution_comparison} shows a comparison between an atmospheric transmission spectrum computed using a 1\,GHz spectral resolution (in red) and a 1\,MHz spectral resolution (in black).  While the higher resolution spectrum shows an increased density of fine-scale absorption features, we see that the 1\,GHz spectral resolution is sufficient to capture the dominant structures expected to be relevant for continuum observations.

\section{Temporal resolution} \label{app:ResolutionInTime}

\begin{figure*}
    \centering
    \includegraphics[width=1.00\textwidth]{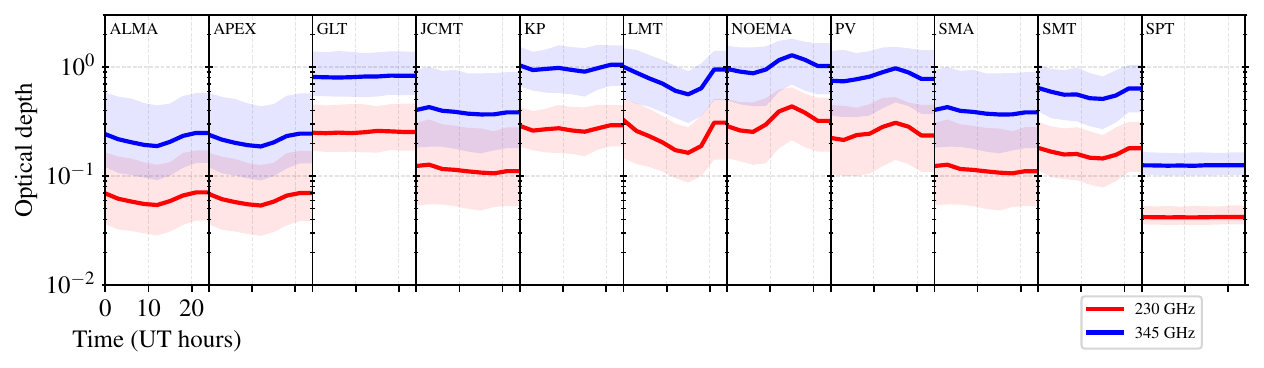}
    \caption{Optical depth versus time of day at observing frequencies of 230\,GHz (in red) and 345\,GHz (in blue), assuming observing conditions appropriate for April at each of the stations currently participating in the EHT array.  For both frequencies, the central line shows the median optical depth, and the shaded region brackets the 68th percentile interval (i.e., spanning the 16th to the 84th percentile values).  All panels share the same horizontal and vertical axis ranges, which are explicitly labeled in the left panel.}\label{fig:daily_weather}
\end{figure*}

\autoref{fig:daily_weather} shows the optical depth versus time of day for each of the sites currently participating in the EHT array, assuming April observing conditions.  The optical depth time series are plotted at the native 3-hour temporal resolution of the MERRA-2 database, prior to carrying out the daily averaging that is used within \ngehtsim.  While some sites (e.g., LMT, NOEMA) exhibit clear differences between the daytime and nighttime optical depths, many of the sites experience only modest variations.  In all cases, the magnitude of the typical day-to-day variation -- captured by the shaded regions in each panel -- is comparable to or greater than the magnitude of the typical daytime-to-nighttime variation.  Weather parameters that have been averaged on a per-day basis thus retain most of the inter-day variation that is relevant for generating realistic synthetic data using \ngehtsim.

\section{Principal component analysis decomposition of atmospheric spectra} \label{app:Decomposition}

Because we generate atmospheric optical depth and brightness temperature spectra with values spaced every 1\,GHz from 0\,THz to 2\,THz, each of our spectra contains $N_{\nu} = 2001$ values.  Given that \ngehtsim requires access to the optical depth and brightness temperature spectra corresponding to every site at every day over a time range that spans more than a decade, we carry out principal component analysis (PCA) decomposition of these spectra to reduce the overall data volume.

Given $N$ spectra $\boldsymbol{t}_i$, we compute a mean spectrum

\begin{equation}
\langle \boldsymbol{t} \rangle = \frac{1}{N} \sum_{i=1}^N \boldsymbol{t}_i
\end{equation}

\noindent and subtract it from each of the individual spectra to generate centered spectra,

\begin{equation}
\boldsymbol{\hat{t}}_i = \boldsymbol{t}_i - \langle \boldsymbol{t} \rangle .
\end{equation}

\noindent We then aggregate these centered spectra into a $N \times N_{\nu}$ matrix $\textbf{X}$, 

\begin{equation}
\textbf{X} = \begin{pmatrix}
\boldsymbol{\hat{t}}_1^{\top} \\
\boldsymbol{\hat{t}}_2^{\top} \\
\vdots \\
\boldsymbol{\hat{t}}_N^{\top}
\end{pmatrix} = \begin{pmatrix}
\hat{t}_1(\nu_1) & \hat{t}_1(\nu_2) & \ldots & \hat{t}_1(\nu_{N_{\nu}}) \\
\hat{t}_2(\nu_1) & \hat{t}_2(\nu_2) & \ldots & \hat{t}_2(\nu_{N_{\nu}}) \\
\vdots & \vdots & \ddots & \vdots \\
\hat{t}_N(\nu_1) & \hat{t}_N(\nu_2) & \ldots & \hat{t}_N(\nu_{N_{\nu}})
\end{pmatrix} ,
\end{equation}

\noindent whose associated $N_{\nu} \times N_{\nu}$ covariance matrix is given by

\begin{equation}
\boldsymbol{\Sigma} = \textbf{X}^{\top} \textbf{X} .
\end{equation}

The orthonormal set of eigenvectors $\boldsymbol{v}_j$ of $\boldsymbol{\Sigma}$ corresponds to the principal components, or ``eigenspectra,'' of our decomposition.  Each spectrum $\boldsymbol{t}_i$ can then be approximately reconstructed using an appropriate linear combination of eigenspectra,

\begin{equation}
\boldsymbol{t}_i \approx \langle \boldsymbol{t} \rangle + \sum_{j=1}^{n} A_{i,j} \boldsymbol{v}_j .
\end{equation}

\noindent The coefficients $A_{i,j}$ are obtained from projection of the centered spectra onto the eigenbasis,

\begin{equation}
A_{i,j} = \boldsymbol{\hat{t}}_i \cdot \boldsymbol{v}_j ,
\end{equation}

\noindent and the number $n$ of eigenspectra to use in the reconstruction depends on the desired trade-off between the fidelity of approximation and the degree of compression.

To ensure a high-fidelity approximation across many orders of magnitude in $\tau$, we perform PCA on the logarithm of the optical depth; our decomposition of $T_b$ uses the linear spectrum.  We use $\sim$1.6 million full spectra -- corresponding to every spatial and temporal point contained in the MERRA-2 database for April 15, 2022 -- to determine the PCA decomposition for both $\tau$ and $T_b$.  We find that the first $n=40$ principal components are typically sufficient to achieve a maximum residual of $\lesssim 1$\,K in $T_b$ and $\lesssim 10^{-2}$ in each of the optical depth, log optical depth, and transmission.  Recording only the coefficients corresponding to these components yields a storage savings of approximately a factor of 50.

\section{Receiver properties} \label{app:Receivers}

\autoref{tab:Receivers} lists the telescope and receiver properties assumed for EHT stations in the simulations carried out in this paper.

\begin{deluxetable*}{lcccccccc}
\tablecolumns{9}
\tablewidth{0pt}
\tablecaption{Receiver properties for EHT stations\label{tab:Receivers}}
\tablehead{& \multicolumn{3}{c}{1.3\,mm receiver} & & \multicolumn{3}{c}{0.87\,mm receiver} \\ \cline{2-4} \cline{6-8}
Station & Frequency & $T_{\text{rx}}$ & $r$ & & Frequency & $T_{\text{rx}}$ & $r$ & References}
\startdata
ALMA  & 211--275\,GHz & 40\,K & 0.01 & & 275--373\,GHz & 75\,K & 0.1 & \citet{Kerr_2004,Mahieu_2012} \\
APEX (2017)  & 211--275\,GHz & 90\,K$^a$ & 0.03 & & \ldots & \ldots & \ldots & \citet{Vassilev_2008} \\
APEX (post-2017)  & 196--281\,GHz & 85\,K$^b$ & 0.03 & & 272--376\,GHz & 120\,K & 0.03 & \citet{Meledin_2022} \\
GLT & 207--235\,GHz & 70\,K & 0.01 & & 275--373\,GHz & 150\,K & 0.1 & \citet{Hasegawa_2017,Han_2018} \\
JCMT (2017) & 215--270\,GHz & 50\,K & 1.25 & & \ldots & \ldots & \ldots & JCMT website$^c$ \\
JCMT (post-2017) & 212--273\,GHz & 60\,K & 0.03 & & 275--373\,GHz & 80\,K & 0.03 & \citet{Mizuno_2020} \\
KP    & 211--275\,GHz & 80\,K$^d$ & 0.03$^{d}$ & & \ldots & \ldots & \ldots & \ldots \\
LMT (2017)   & 209--233\,GHz & 130\,K$^e$ & 1     & & \ldots & \ldots & \ldots & \citetalias{M87Paper3} \\
LMT (post-2017)   & 210--280\,GHz & 70\,K & 0.03$^f$     & & \ldots & \ldots & \ldots & \citet{Bustamante_2023} \\
NOEMA & 200--276\,GHz & 80\,K & 0.1 & & 275--373\,GHz & 150\,K & 0.1 & \citet{Chenu_2016} \\
PV  & 200--267\,GHz & 60\,K & 0.03 & & 260--360\,GHz & 85\,K & 0.03 & \citet{Carter_2012} \\
SMA   & 194--281\,GHz & 70\,K & 1     & & 258--408\,GHz & 130\,K & 1 & \citet{Wilner_1998} \\
SMT   & 205--280\,GHz & 80\,K & 0.03 & & 325--370\,GHz & 150\,K & 1 & SMT website$^g$ \\
SPT   & 212--230\,GHz & 40\,K & 0.03 & & \ldots & \ldots & \ldots & \citet{Kim_2018} \\
\enddata
\tablecomments{Receiver properties assumed for the simulations carried out in \autoref{sec:EHT2017}, \autoref{sec:EHT345}, and \autoref{sec:EHTmultifreq}.  The APEX, JCMT, and LMT sites received upgrades to their equipment after the initial 2017 EHT observations, so there are two sets of receiver properties provided for each of these sites. \\
$^a$The receiver temperature measured for the APEX-1 receiver during commissioning was $\sim$120\,K \citep{Vassilev_2008}, but receiver improvements over time reduced this value to $\sim$90\,K by 2017 (see \url{https://www.apex-telescope.org/heterodyne/shfi/het230/characteristics/index.php}).\\
$^b$See \url{http://www.apex-telescope.org/ns/nflash/}.\\
$^c$See \url{https://www.eaobservatory.org/jcmt/instrumentation/heterodyne/rxa/}.\\
$^d$Assumed to match SMT.\\
$^e$The 2017 LMT receiver temperature has been estimated from the system temperature measurements in \citet{M87Paper3}, assuming an atmospheric temperature of 273\,K and optical depth of 0.2.\\
$^f$See \url{http://lmtgtm.org/single-pixel-1mm-receiver/}.\\
$^g$See \url{https://aro.as.arizona.edu/?q=facilities/submillimeter-telescope}. Receiver temperatures for the SMT have been estimated from the provided system temperature measurements assuming an atmospheric temperature of 273\,K and optical depth of 0.2.}
\end{deluxetable*}

\section{Site information} \label{app:SiteTable}

\autoref{tab:Sites} lists the locations, elevations, effective dish diameters, and surface accuracies for of a number of existing and near-future sites for which atmospheric information is available in \ngehtsim.

\begin{deluxetable*}{lcccccc}
\tablecolumns{7}
\tablewidth{0pt}
\tablecaption{Information for sites available in \ngehtsim \label{tab:Sites}}
\tablehead{Site code & Location & Latitude & Longitude & Elevation (m) & $D$ (m) & $\sigma_{\text{RMS}}$ ($\mu$m)}
\startdata
ALMA  & Atacama, Chile & $-23.032$ & \hphantom{0}$-67.755$ & 5040 & 75 & 25 \\
AMT & Gamsberg, Namibia & $-23.339$ & \hphantom{$-$0}16.229 & 2340 & 15 & 25 \\
APEX & Atacama, Chile & $-23.005$ & \hphantom{0}$-67.759$ & 5060 & 12 & 25 \\
ATCA & New South Wales, Australia & $-30.313$ & \hphantom{$-$}$149.564$ & 210 & 54 & 200 \\
EFF	&	Cologne, Germany	& \hphantom{$-$}50.525	& \hphantom{$-$00}6.884 & 390 & 100 & 550 \\
GBT   & West Virginia, US & \hphantom{$-$}38.434 & \hphantom{0}$-79.840$ & 810 & 100 & 260 \\
GLT & Pituffik Space Base, Greenland & \hphantom{$-$}76.535 & \hphantom{0}$-68.686$ & 70 & 12 & 50 \\
HAY & Massachusetts, US & \hphantom{$-$}42.624 & \hphantom{0}$-71.489$ & 110 & 37 & 85 \\
JCMT & Mauna Kea, Hawaii & \hphantom{$-$}19.823 & $-155.477$ & 4070 & 15 & 24 \\
KP & Arizona, US & \hphantom{$-$}31.953 & $-111.615$ & 1930 & 12 & 16 \\
KVNPC & Pyeongchang, South Korea & \hphantom{$-$}37.534 & \hphantom{$-$}128.450 & 500 & 21 & 70 \\
KVNTN & Tamna, South Korea & \hphantom{$-$}33.289 & \hphantom{$-$}126.460 & 410 & 21 & 70 \\
KVNUS & Ulsan, South Korea & \hphantom{$-$}35.546 & \hphantom{$-$}129.249 & 130 & 21 & 70 \\
KVNYS & Yonsei, South Korea & \hphantom{$-$}37.565 & \hphantom{$-$}126.941 & 90 & 21 & 70 \\
LLA & Salta, Argentina & $-24.192$ & \hphantom{0}$-66.475$ & 4780 & 12 & 25 \\
LMT & Sierra Negra, Mexico & \hphantom{$-$}18.986 & \hphantom{0}$-97.315$ & 4620 & 50 & 80 \\
MET & Uusimaa, Finland & \hphantom{$-$}60.218 & \hphantom{$-$0}24.393 & 50 & 13.7 & 100 \\
NOB & Nagano Prefecture, Japan & \hphantom{$-$}35.944 & \hphantom{$-$}138.472 & 1370 & 45 & 100 \\
NOEMA & Plateau de Bure, France & \hphantom{$-$}44.634 & \hphantom{$-$00}5.907 & 2550 & 50 & 35 \\
ONS & Halland County, Sweden & \hphantom{$-$}57.396 & \hphantom{$-$0}11.926 & 30 & 20 & 128 \\
OVRO & California, US & \hphantom{$-$}37.231 & $-118.282$ & 1210 & 10.4 & 40 \\
PV & Sierra Nevada, Spain & \hphantom{$-$}37.066 & \hphantom{00}$-3.393$ & 2860 & 30 & 55 \\
SMA & Mauna Kea, Hawaii & \hphantom{$-$}19.824 & $-155.478$ & 4070 & 15 & 20 \\
SMT & Arizona, US & \hphantom{$-$}32.702 & $-109.891$ & 3170 & 10 & 15 \\
SPT & South Pole, Antarctica & $-90.000$ & \hphantom{$-$00}0.000 & 2820 & 10 & 25 \\
VLA &   New Mexico, US  & \hphantom{$-$}34.079  & $-107.618$ &   2120 & 130 & 420 \\
VLBBR &   Washington, US  & \hphantom{$-$}48.131  & $-119.683$ &   260 & 25 & 320 \\
VLBFD &   Texas, US  & \hphantom{$-$}30.635  & $-103.945$ &   1610 & 25 & 320 \\
VLBHN &   New Hampshire, US  & \hphantom{$-$}42.934  & \hphantom{0}$-71.987$ &   310 & 25 & 320 \\
VLBKP &   Arizona, US  & \hphantom{$-$}31.956  & $-111.612$ &   1920 & 25 & 320 \\
VLBLA &   New Mexico, US  & \hphantom{$-$}35.775  & $-106.246$ &   1970 & 25 & 320 \\
VLBMK &   Mauna Kea, Hawaii  & \hphantom{$-$}19.802  & $-155.456$ &   3730 & 25 & 320 \\
VLBNL &   Iowa, US  & \hphantom{$-$}41.771  & \hphantom{0}$-91.574$ &   240 & 25 & 320 \\
VLBOV &   California, US  & \hphantom{$-$}37.232  & $-118.277$ &   1210 & 25 & 320 \\
VLBPT &   New Mexico, US  & \hphantom{$-$}34.301  & $-108.119$ &   2370 & 25 & 320 \\
VLBSC &   St. Croix, US Virgin Islands  & \hphantom{$-$}17.757  & \hphantom{0}$-64.584$ &   10 & 25 & 320 \\
YEB &   Castilla–La Mancha, Spain  & \hphantom{$-$}40.523  & \hphantom{00}$-3.088$ &   920 & 40 & 150 \\
\enddata
\tablecomments{Location, elevation, effective dish diameter ($D$), and surface accuracy ($\sigma_{\text{RMS}}$) for existing and near-future sites available in \ngehtsim.}
\end{deluxetable*}

\section{Geometric source structure models for \m87 and \sgra} \label{app:SourceModels}

\begin{figure*}
    \centering
    \includegraphics[width=1.00\textwidth]{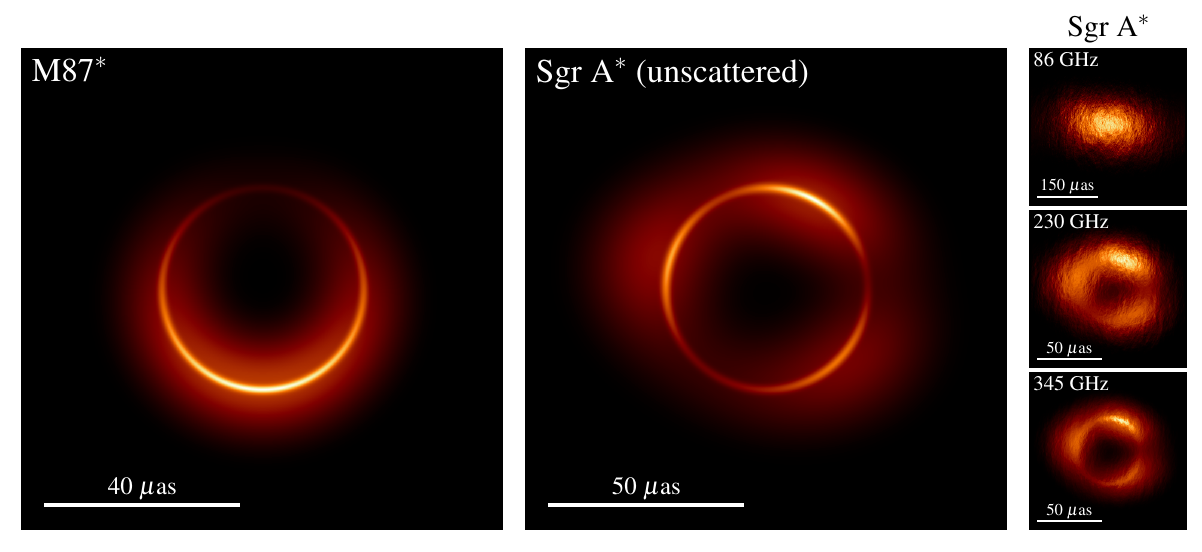}
    \caption{Source models of \m87 (left panel) and \sgra (rightmost four panels) used for the simulations carried out in this paper.  The source structure for \m87 is assumed to be independent of frequency, while the \sgra source structure evolves substantially with frequency because of the effects of interstellar scattering.  The second panel from the left shows the underlying unscattered \sgra source structure, which we assume to be independent of frequency.  The three smaller panels on the right show the corresponding on-sky \sgra source structure after applying the effects of interstellar scattering at three different observing frequencies; from top to bottom, the three observing frequencies are 86\,GHz, 230\,GHz, and 345\,GHz.} \label{fig:source_models}
\end{figure*}

A key property determining the detectability of a source on interferometric baselines is the source's emission structure.  For the simulations carried out in this paper, we use tailored geometric source structure models to mimic the horizon-scale appearances of \m87 and \sgra at the (sub)millimeter wavelengths of interest.  Our basic model building block is the so-called ``m-ring'' geometric model, introduced in \cite{Johnson_2020} and used extensively in \cite{SgrA_Paper4} to model the observed structure of \sgra.  While purely phenomenological, the m-ring model has the benefit of capturing the gross features observed by the EHT to be present in the horizon-scale emission structure around \m87 and \sgra using a small number of parameters.

An infinitesimally thin m-ring can be parameterized in terms of its total flux density $S_0$, its diameter $D$, an elongation parameter $s$, an orientation angle $\psi$, and a series of complex Fourier coefficients $\beta_k$ that describe the azimuthal intensity distribution.  In terms of these parameters, we can express the image intensity $I$ in modified polar coordinates $(r,\phi)$ as

\begin{equation}
I_{\text{ring}}(r,\phi) = \frac{S_0}{\pi D s} \delta\left( r - \frac{D}{2} \right) \sum_{k=-m}^m \beta_k e^{i k \phi} . \label{eqn:DeltaMRing}
\end{equation}

\noindent Here, $m$ is an integer that sets the order of the azimuthal expansion, $\beta_k = \beta_k^*$ for all $k$ because the image has real-valued intensities, and we fix $\beta_0 = 1$ so that $S_0$ sets the total flux density.  The modified polar coordinates are related to modified Cartesian coordinates $(x',y')$ via

\begin{eqnarray}
r^2 & = & x'^2 + y'^2 \nonumber \\
\tan\left( \phi \right) & = & \frac{y'}{x'} ,
\end{eqnarray}

\noindent which in turn are related to the standard image-plane Cartesian coordinates $(x,y)$ via the stretching transformation

\small
\begin{equation}
\begin{pmatrix}
x' \\
y'
\end{pmatrix} = \begin{pmatrix}
\cos^2(\psi) + \frac{\sin^2(\psi)}{s} & \cos(\psi) \sin(\psi) \left( \frac{1}{s} - 1 \right) \\
\cos(\psi) \sin(\psi) \left( \frac{1}{s} - 1 \right) & \frac{\cos^2(\psi)}{s} + \sin^2(\psi)
\end{pmatrix} \begin{pmatrix}
x \\
y
\end{pmatrix} .
\end{equation}
\normalsize

\noindent To produce an m-ring with a finite thickness, we then convolve \autoref{eqn:DeltaMRing} with a circular Gaussian having FWHM $\alpha$,

\begin{equation}
I_{\text{m-ring}}(r,\phi) = \frac{4 \ln(2)}{\pi \alpha^2} \exp\left( - \frac{4 \ln(2) r^2}{\alpha^2} \right) * I_{\text{ring}}(r,\phi) . \label{eqn:MRing}
\end{equation}

\noindent We use \autoref{eqn:MRing} to produce mock images of both \m87 and \sgra for the simulations carried out in this paper.

We model \m87 using a sum of two concentric m-rings.  This model choice is intended to capture the expected structure corresponding to the first two images of the multiply-lensed emission, with a thicker m-ring representing the direct emission and a thinner m-ring representing emission from the first-order lensed image or ``photon ring'' \citep{Johnson_2020}.  Both m-rings have order $m=1$, corresponding to a single $\sim$dipolar asymmetry in the otherwise ring-like emission structure \citep[see][]{M87Paper4,M87Paper6}.  We set the total flux density to be $S_0 = 0.6$\,Jy, with 90\% of the flux density in the direct emission and 10\% of the flux density in the photon ring.  The ring diameter is set to be $D = 42$\,\uas for both m-rings, and the ring width is $\alpha = 15$\,\uas for the direct emission and $\alpha = 1.5$\,\uas for the photon ring.  The single Fourier coefficient is set to be $\beta_1 = -0.4$ for both m-rings.  The left panel of \autoref{fig:source_models} shows the \m87 image structure resulting from this model; for the simulations carried out in this paper, we assume that the this source structure is independent of frequency.

We similarly model \sgra using a sum of two concentric m-rings.  Both m-rings have order $m=4$, permitting sufficient flexibility to capture the more complex azimuthal intensity structure observed towards \sgra \citep[see][]{SgrA_Paper2,SgrA_Paper3,SgrA_Paper4,Wielgus_2022}.  We set the total flux density to be $S_0 = 2.5$\,Jy, again apportioned such that 90\% of the flux density is in the direct emission and 10\% of the flux density is in the photon ring.  The ring diameter is set to be $D = 52$\,\uas for both m-rings, and the ring width is $\alpha = 20$\,\uas for the direct emission and $\alpha = 2$\,\uas for the photon ring.  The four Fourier coefficients are set to be the same for both m-rings, taking on values of 

\begin{eqnarray*}
\beta_1 & = & 0.15 \\
\beta_2 & = & 0.06 \exp\left( -2.3 \pi i \right) \\
\beta_3 & = & 0.25 \exp\left( 0.5 \pi i \right) \\
\beta_4 & = & 0.13 \exp\left( -1.4 \pi i \right) .
\end{eqnarray*}

\noindent The stretch parameter is set to be $s = 1.2$ for the direct emission and $s = 1$ for the photon ring, and the orientation parameter is $\psi = (\pi/2) - 0.2$ for both m-rings.  We apply frequency-dependent interstellar scattering effects to the images using the ``stochastic optics'' package \citep{Johnson_2016} as implemented within \texttt{ehtim}, which uses the scattering screen parameters measured by \cite{Johnson_2018}.  The corresponding images of \sgra are shown in the right four panels of \autoref{fig:source_models}; the large panel shows the unscattered image structure (which we assume to be the same at all frequencies), and the three smaller rightmost panels show the scattered images at observing frequencies of 86\,GHz, 230\,GHz, and 345\,GHz.

\section{Atmospheric phase variations} \label{app:AtmosphericPhase}

Variations in the amount of atmospheric water vapor cause corresponding variations in the index of refraction, which then serve as the primary source of rapid phase fluctuations at (sub)millimeter wavelengths \citep{Thompson_2017}.  The standard model for these phase fluctuations is that they can be described by a phase screen obeying the so-called ``Taylor hypothesis'' \citep{Taylor_1938}, which assumes that the turbulent structures giving rise to the refractive variations in the atmosphere are ``frozen'' over the observational timescales of interest.  Typical practice is to describe the distribution of phase turbulence in terms of a ``structure function'' \citep[e.g.,][]{Treuhaft_1987,Asaki_2007,Natarajan_2022}

\begin{equation}
D(r) = \langle \left[ \phi\left( \boldsymbol{x} + \boldsymbol{r} \right) - \phi\left( \boldsymbol{x} \right) \right]^2 \rangle ,
\end{equation}

\noindent where $\boldsymbol{x}$ and $\boldsymbol{r}$ are transverse position vectors in the screen and the angle brackets denote an ensemble average.  The structure function provides the mean squared phase difference between two locations on the screen separated by $\boldsymbol{r}$, and we have implicitly assumed isotropy by casting the structure function in terms of the magnitude of this separation, $r = |\boldsymbol{r}|$.

Given the frozen screen assumption, the spatial phase variations can be related to temporal phase variations through the transformation $r \rightarrow v t$, where $v \approx 10$\,m\,s$^{-1}$ is the transverse velocity of the phase screen and $t$ is time.  For a wide range of spatial scales, the structure function is then well-described by a power law in $t$ \citep{Kolmogorov_1941,Dravskikh_1979,Coulman_1985},

\begin{equation}
D(t,\lambda) = \left( \frac{\text{1\,mm}}{\lambda} \right)^2 \left( \frac{t}{t_{c,\text{1\,mm}}} \right)^{\alpha} . \label{eqn:SFAtmosphere}
\end{equation}

\noindent where $t_{c,\text{1\,mm}}$ is the coherence timescale on which the phase variance is equal to 1 radian at an observing wavelength of $\lambda = 1$\,mm.  The value of $\alpha$ depends on the relative size of the observing wavelength and the length scales governing the turbulent medium; for wavelengths shorter than the vertical extent of the turbulent layer ($\sim$1\,km; \citealt{Carilli_1999}) but longer than the dissipative scale ($\sim$1\,mm; \citealt{Stotskii_1973}), $\alpha$ takes on the value of ${\sim}5/3$ appropriate for 3D Kolmogorov turbulence \citep{Thompson_2017}.  We will adopt the value $\alpha = 5/3$ for the remainder of this paper.

A typical 1\,mm coherence timescale at a centimeter-wavelength site like the VLA is $t_{c,\text{1\,mm}} \approx 10$ seconds \citep{Beasley_1995}, while for the millimeter-wavelength sites used in the EHT array the coherence timescale can be longer by a factor of several \citep{M87Paper2}.

\subsection{Phase tracking: sensitivity and integration time considerations} \label{sec:PhaseTracking}

Calibrating the phase measurements in interferometric observations requires being able to accurately track the phase fluctuations induced by geometric, instrumental, and atmospheric effects, which in turn requires a certain sensitivity to be achieved in the phase measurements.  Regardless of the specific precision requirements for any particular experiment, given some fixed observing conditions there will always be some integration time below which the signal-to-noise ratio (SNR) $\rho$ is insufficient to provide the target precision.  We would thus like to determine the integration time $t_{\text{int}}$ that is necessary to achieve a desired threshold SNR $\rho_{\text{thresh}}$ in our phase measurement.  We assume that once this threshold SNR is achieved, the phases can be calibrated sufficiently well to integrate for arbitrarily long times, assuming that the source visibility is not evolving appreciably during the integration period (or that it can be modeled well).

Denoting the thermal phase noise on an integration time of $t_0$ by $\sigma_{\text{th},0}$, the thermal phase noise $\sigma_{\text{th}}(t_{\text{int}})$ on an integration time $t_{\text{int}}$ is given by

\begin{equation}
\sigma_{\text{th}}(t_{\text{int}}) = \sigma_{\text{th},0} \left( \frac{t_0}{t_{\text{int}}} \right)^{1/2} . \label{eqn:ThermalPhaseNoise}
\end{equation}

\noindent Here, we note that the reference thermal noise $\sigma_{\text{th},0}$ does not depend on any characteristics of the atmospheric turbulence; it is purely a description of the measurement sensitivity, which will typically be determined by properties such as the collecting areas of the telescopes, the level of receiver noise, the integrated frequency bandwidth, and the brightness of the observed source.

The existence of intrinsic phase fluctuations on all timescales means that any amount of averaging will miss some phase variations, primarily those associated with timescales shorter than the averaging time $t_{\text{int}}$.  I.e., there is an additional source of phase noise that arises during the averaging process itself, which is associated with the intrinsic evolution of the phase with time over the duration of the averaging period.  For a signal with structure function given by \autoref{eqn:SFAtmosphere}, the residual phase variance $\sigma_{\text{res},i}^2$ contributed by station $i$ after boxcar averaging over an interval $t_{\text{int}}$ is given by \citep{Blackburn_2019}

\begin{equation}
\sigma_{\text{res},i}^2 = \frac{2^{-\alpha} (2 + \alpha - 2^{\alpha})}{(1 + \alpha) (2 + \alpha)} \left( \frac{t_{\text{int}}}{t_{c,i}} \right)^{\alpha} \equiv K \left( \frac{t_{\text{int}}}{t_{c,i}} \right)^{\alpha} , \label{eqn:ResidualPhaseNoise}
\end{equation}

\noindent where $t_{c,i}$ is the coherence time at station $i$ and $K$ is a constant that depends only on $\alpha$.  For $\alpha = 5/3$, $K \approx 0.016$.  It is convenient to define an effective coherence time $t_c$,

\begin{equation}
t_{c} \equiv \left( \frac{1}{t_{c,i}^{\alpha}} + \frac{1}{t_{c,j}^{\alpha}} \right)^{-1/\alpha} ,
\end{equation}

\noindent such that 

\begin{equation}
\sigma_{\text{res},i}^2 + \sigma_{\text{res},j}^2 = K \left( \frac{t_{\text{int}}}{t_c} \right)^{\alpha} . \label{eqn:ResidualPhaseNoiseBaseline}
\end{equation}

\noindent The quantity $t_c$ provides a notion of the coherence time for the baseline connecting stations $i$ and $j$, rather than for each station individually.  If $t_{c,i} = t_{c,j}$ and $\alpha = 5/3$, then the effective coherence time is shortened by a factor of $\sim$0.66 relative to that at either station.

The total phase noise $\sigma_{\text{tot}}$ incurred after averaging on a timescale $t_{\text{int}}$ will then be determined by a combination of the thermal (\autoref{eqn:ThermalPhaseNoise}) and residual (\autoref{eqn:ResidualPhaseNoiseBaseline}) phase noises,

\begin{equation}
\sigma_{\text{tot}}(t_{\text{int}}) = \sqrt{\sigma_{\text{th}}^2 + \sigma_{\text{res},i}^2 + \sigma_{\text{res},j}^2} . \label{eqn:TotalPhaseNoise}
\end{equation}

\begin{figure}
    \centering
    \includegraphics[width=1.00\columnwidth]{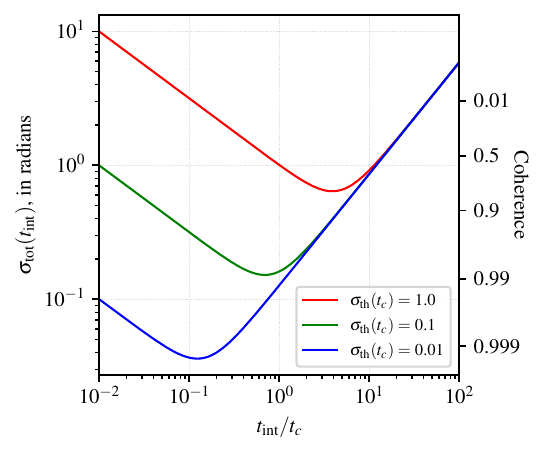}
    \caption{Total phase noise (\autoref{eqn:TotalPhaseNoise}) as a function of integration time.  Three different baseline sensitivity values -- characterized by their thermal phase noise levels on a coherence time -- are plotted as different colored lines and labeled accordingly.  The righthand vertical axis indicates the phase coherence corresponding to a particular total phase noise value, where the coherence is given by $\exp(-\sigma_{\text{tot}}^2/2)$.} \label{fig:sigma_vs_t}
\end{figure}

\noindent For stations that are geographically separated and thus observing through different atmospheric phase screens -- as is typically the case for VLBI -- we expect $\sigma_{\text{res},i}$ and $\sigma_{\text{res},j}$ to be independent, such that their variances simply add in \autoref{eqn:TotalPhaseNoise}.  \autoref{fig:sigma_vs_t} shows $\sigma_{\text{tot}}(t_{\text{int}})$ for several different choices of baseline sensitivity.

Given fixed values of $\alpha$, $\sigma_{\text{th},0}$, $t_0$, and $t_c$, there is a value of $t_{\text{int}}$ that minimizes $\sigma_{\text{tot}}$.  This minimum value of $\sigma_{\text{tot}}$ is given by

\begin{eqnarray}
\sigma_{\text{min}} & = & \sigma_{\text{th},c}^{\alpha/(1+\alpha)} \sqrt{\left( \alpha^{1/(1+\alpha)} + \alpha^{-\alpha/(1+\alpha)} \right) K^{1/(1+\alpha)}} \nonumber \\
& \approx & 0.64 \sigma_{\text{th},c}^{5/8} , \label{eqn:MinPhaseNoise}
\end{eqnarray}

\noindent where for convenience we have taken $t_0 = t_c$ and thus relabeled $\sigma_{\text{th},0} \rightarrow \sigma_{\text{th},c} = \sigma_{\text{th}}(t_c)$, and in the second line we have explicitly plugged in $\alpha = 5/3$.  The corresponding integration time required to achieve $\sigma_{\text{min}}$ is given by

\begin{eqnarray}
\frac{t_{\text{max}}}{t_c} & = & \left( \frac{\sigma_{\text{th},c}^2}{\alpha K} \right)^{1/(1+\alpha)} \nonumber \\
& \approx & 3.91 \sigma_{\text{th},c}^{3/4} , \label{eqn:OptimalIntegration}
\end{eqnarray}

\noindent where in the second line we have again explicitly plugged in $\alpha = 5/3$.  $t_{\text{max}}$ represents the longest integration time that will typically be useful for a given value of $\sigma_{\text{th},c}$; integrating for longer than $t_{\text{max}}$ will result in increased total phase noise.  If the minimum phase noise that a baseline can achieve never falls below some threshold value $\sigma_{\text{thresh}}$, then the phase on that baseline cannot be accurately tracked and the visibilities on that baseline are effectively undetected.  The values of both $t_{\text{max}}$ and $\sigma_{\text{min}}$ are determined by the baseline sensitivity ($\sigma_{\text{th}}$) and effective coherence time ($t_c$), as shown in \autoref{fig:sigma_vs_t}.

The rule of thumb we follow for the simulations in this paper is that the thermal SNR on a baseline must be at least $\rho_{\text{thresh}} \geq 5$ on an integration time of $t_{\text{int}} = t_c/3$ (see \autoref{sec:VisDetection}).  Per \autoref{eqn:ThermalPhaseNoise}, this condition corresponds to $\sigma_{\text{th},c} \approx 0.12$ radians, which in turn implies that $\sigma_{\text{min}} \approx 0.17$ radians (\autoref{eqn:MinPhaseNoise}) and $t_{\text{max}} \approx 0.77 t_c$ (\autoref{eqn:OptimalIntegration}).  I.e., requiring that the thermal SNR be at least 5 on an integration time of $t_c/3$ is sufficient to ensure that the achievable total phase noise is better than $\sim$0.2 radians.  For comparison, typical SNR values for EHT observations of \m87 are ${\sim}0.3$--10 on non-ALMA baselines within a $t_c/3 \approx 10$\,s integration time \citep{M87Paper3}; by our rule-of-thumb criterion, the visibility phases on these baselines cannot always be tracked without some alternative source of phase stabilization.

\subsection{Phase tracking: multi-frequency} \label{sec:PhaseTrackingMultiFreq}

The tropospheric phase fluctuations that dominate the atmospheric component of the phase variations at high radio frequencies are largely non-dispersive, meaning that the magnitude of the phase variations increases linearly with frequency (see \autoref{eqn:SFAtmosphere}).  This behavior means that if the atmospheric phase variations can be tracked at one frequency, then they can then be used to calibrate concurrent measurements made at another frequency by scaling the phase solutions by the frequency ratio.  This calibration technique -- known as frequency phase transfer \citep[FPT;][]{Rioja_2020} -- leverages the relative ease of observations at lower frequencies to bolster or even enable observations at higher frequencies\footnote{Note that FPT is not restricted to transferring phase from low frequencies to high frequencies; in principle, the phase could be tracked at the higher frequency and then transferred to the lower frequency.  But in practice, it is almost always the case that the direction of most useful phase transfer goes from lower to higher frequency.}.  FPT-based calibration provides a promising avenue for pushing millimeter VLBI to higher frequencies and fainter flux densities \citep[e.g.,][]{Issaoun_2023,Rioja_2023}, and it is emulated within \ngehtsim.

Let us use $\nu_{\ell}$ and $\nu_h$ to denote the low (reference) and high (target) frequencies, respectively.  We will define

\begin{equation}
R = \frac{\nu_h}{\nu_{\ell}} = \left( \frac{t_{c,\ell}}{t_{c,h}} \right)^{\alpha/2}
\end{equation}

\noindent to be the frequency ratio.  The transfer of phase from the reference frequency to the target frequency involves a multiplication by the frequency ratio $R$, which also inflates the noise in the phase measurement.  The resulting additional phase noise transferred to the target frequency from the reference frequency is then given by

\begin{equation}
\sigma_{\text{FPT}} = R \sqrt{ \sigma_{\text{th},\ell}^2 + \sigma_{\text{res},\ell,1}^2 + \sigma_{\text{res},\ell,2}^2 } , \label{eqn:FPTNoise}
\end{equation}

\noindent where $\sigma_{\text{th},\ell}$ is the thermal noise at the reference frequency and $\sigma_{\text{res},\ell,i}$ is the residual phase noise at the reference frequency from station $i$.

Up to a constant scaling factor of $R$, the excess phase noise imparted during FPT (\autoref{eqn:FPTNoise}) is the same as the total phase noise of the reference frequency phase signal (\autoref{eqn:TotalPhaseNoise}).  The optimal integration time $t_{\text{max}}$ that minimizes $\sigma_{\text{FPT}}$ will thus remain the same as in \autoref{eqn:OptimalIntegration}, while the effective noise in the transferred phase -- i.e., ignoring the thermal noise at the target frequency itself -- will be increased by a factor $R$ relative to that in \autoref{eqn:MinPhaseNoise}.  So to achieve any given threshold noise level level $\sigma_{\text{thresh}}$ in the phase transferred to the target frequency requires a factor $R$ lower phase noise at the reference frequency.

This scaling also means that FPT is only useful in practice -- i.e., it will only provide improved phase tracking over that obtained through single-frequency observations -- if the phase measurement at the reference frequency is sufficiently more sensitive than the phase measurement at the target frequency.  The thresholding requirement is that the phase measurement at the reference frequency must satisfy $\sigma_{\text{th},0,\ell} < \sigma_{\text{th},0,h} / R$; i.e., the thermal phase noise at the reference frequency (within any given integration time) must be a factor of $R$ smaller than the corresponding thermal phase noise at the target frequency.  We expect that lower-frequency observations will typically be more sensitive than those at higher frequencies, both because characteristic system noises tend to increase with frequency as well as because sources tend to have higher correlated flux densities on shorter dimensionless baseline lengths.  However, the answer to the question of whether any particular reference-target frequency pair is suitable for FPT will ultimately be both source- and telescope-dependent.

\end{document}